\documentclass[pdflatex,sn-mathphys-num]{sn-jnl}

\usepackage{graphicx}%
\usepackage{multirow}%
\usepackage{amsmath,amssymb,amsfonts}%
\usepackage{amsthm}%
\usepackage{mathrsfs}%
\usepackage[title]{appendix}%
\usepackage[HTML]{xcolor}%
\usepackage{textcomp}%
\usepackage{manyfoot}%
\usepackage{booktabs}%
\usepackage{algorithm}%
\usepackage{algorithmicx}%
\usepackage{algpseudocode}%
\usepackage{listings}%
\usepackage{csquotes}
\usepackage{geometry}
\makeatletter
\@twosidefalse
\@mparswitchfalse
\makeatother
\theoremstyle{thmstyleone}%

\theoremstyle{thmstyletwo}%

\theoremstyle{thmstylethree}%

\definecolor{lavender}{HTML}{C1A4FF}
\definecolor{darkpurple}{HTML}{773299}
\definecolor{peach}{HTML}{ffb47b}
\definecolor{maroon}{HTML}{bf005d}
\definecolor{limegreen}{HTML}{55d24f}
\definecolor{darkblue}{HTML}{325384}
\definecolor{royalblue}{HTML}{0286f0}

\newcommand{\patch}[1]{\fcolorbox{#1}{#1}{\textcolor{#1}{\rule{0.5em}{0.5em}}}}

\begin{document}

\title[The collective dynamics of online harassment]{The collective dynamics of online harassment}

\author*[1]{\fnm{Benjamin} Freixas \sur{Emery}}\email{ben.emery@colorado.edu}

\author[1]{\fnm{Brian} C. \sur{Keegan}}\email{brian.keegan@colorado.edu}

\affil*[1]{\orgdiv{Department of Information Science}, \orgname{University of Colorado Boulder}, \orgaddress{\street{1045 18th Street}, \city{Boulder}, \postcode{80309}, \state{CO}, \country{USA}}}


\abstract{Fringe online message boards are often studied in the context of the extreme ideology that they produce. So far, however, not much of this research has focused on direct real-world harm in the all-too-common form of collective harassment. We directly analyze the complex dynamics of Kiwi Farms, an online message board dedicated largely to the harassment of individuals from vulnerable communities. We conduct exploratory analyses of the hyperlink structure of the platform and linguistic changes over time, and prospective modeling of thread size. After establishing this broader picture of the complex traits of the system, we observe the temporal evolution of community-specific vocabulary, finding that the community's framing of their harassment targets persistently evokes more danger in the early 2020s than the late 2010s. We lastly find that early thread-growth behavior is predictive of longer-term thread virality. We discuss the implications for broader understanding of toxic online behavior and threat assessment, and make the case for studying fringe platforms as complex systems with significant societal impact.}


\maketitle

\section{Introduction}\label{sec:introduction}

Extremist online activity has had increasingly profound effects on modern society, with more people directly or peripherally impacted by the reach of hate groups each day. The daily impact of extremism comes in many forms: mass shooters cite Infowars as their main source of information, people call in bomb threats at schools after its teachers are featured on LibsOfTikTok, and people vote for violent policy after being flooded with rhetoric blaming the most vulnerable in society for their problems. Despite their point of view now being largely reflected in the official policy of the United States government, online extremists do not have any centralized leadership. Some sections of the online far right may even critically disagree with each other. Yet, they are still remarkably effective at targeting specific individuals or communities and affecting policy. The apparent contradiction between these attributes is resolved when viewed through the lens of complex systems. Considering online extremism as a complex sociotechnical system involves focusing on the aggregate emergent behavior of the collection of interactions between the components of the system. 

The present study focuses on the online forum Kiwi Farms: a message board designed and operated specifically for crowdsourcing targeted harassment of minorities, women, LGBTQ people, neurodivergent people, feminists, journalists, internet celebrities, and video game hobbyists \cite{vuNoEasyWay2024}. Using network analysis, natural language processing, and nonlinear growth modeling, we conduct mesoscale analyses to investigate the self-organized structure of the platform, changes in community behavior over time, and early signs of long-term thread salience. In Section \ref{sec:case-study}, we detail how Kiwi Farms is a hub of harassment and terror for the people and communities it chooses to target. Our research questions center on our measurement of this platform's behavior. 

\begin{quote} 
\begin{itemize}
    \item[\textbf{RQ1:}] What mesoscale structures and topical relationships are revealed by the aggregate thread-hyperlink network of Kiwi Farms? 
    \item[\textbf{RQ2:}] What semantic changes surrounding discussion of harassment targets take place over the late 2010s and early 2020s prior to mainstream awareness of this platform's threats to broader society?
    \item[\textbf{RQ3:}] Is long-term size of a thread related to the rate of replies in its first few days, allowing for risk assessment for the targets established in those threads?

\end{itemize}
\end{quote}

\subsection{Related work}\label{sec:related-work}

Online extremism has been a subject of interest for quantitative researchers for nearly two decades, although the constant reintroduction of new platforms and shifting platform popularity has kept few specific discoveries from remaining relevant. Still, the landscape of existing research allows us to at least discuss possible generalities and useful frameworks for studying extremism. For instance, Scrivens et al. demonstrate the efficacy of applying the career criminal framework to modeling extremist posting \cite{scrivensMeasuringEvolutionRadical2020}. Ng et al. use a combination of link frequency and natural language processing to uncover differences in information source but similarities in narrative between Parler and Twitter during the January 6th riots \cite{ngCrossplatformInformationSpread2022}. Cinelli et al. find that discourse between opposing groups increases toxicity of the text \cite{cinelliDynamicsOnlineHate2021}. Hine et al. focus on 4chan, finding that its user-base is global and that its content consists overwhelmingly of hate speech and YouTube links \cite{hineKekCucksGod2017}. There has also been research on misinformation mitigation measures, such as work from McCabe et al. that confirmed the efficacy of deplatforming on Twitter in early 2021 \cite{mccabePostJanuary6thDeplatforming2024}.

In this study, we examine the hyperlink network stemming from a social media platform that harbors hate and extremism. Previous work with link-sharing data has characterized how these link dynamics are proxies for online extremism activity and its influence in mainstream spaces, and has observed how these networks harden in response to major political events and in anticipation of collective action \cite{zhengAdaptiveLinkDynamics2024,vermaHowUSPresidential2024}. 

We also examine the changes in language over time, which has been studied broadly, and for which researchers have developed specific tools \cite{soniAbolitionistNetworksModeling2021,hamiltonDiachronicWordEmbeddings2016,kutuzovDiachronicWordEmbeddings2018}. Semantic change has also been studied in a variety of extremist communities \cite{gothardIncelLexiconDeciphering2021,tornbergWhitePowerEcho2024,tornbergWhiteSupremacistsAnonymous2025}. 

The final portion of this work involves examining the relationship between the early behavior of a thread's growth and its long-term size. This sort of short-long term comparison has been done for view counts on Twitter \cite{pfefferHalfLifeTweet2023}, and for Wikipedia's editor tenures \cite{pancieraWikipediansAreBorn2009}. 

\begin{figure}[h]
    \centering 
    \includegraphics[width=.9\textwidth]{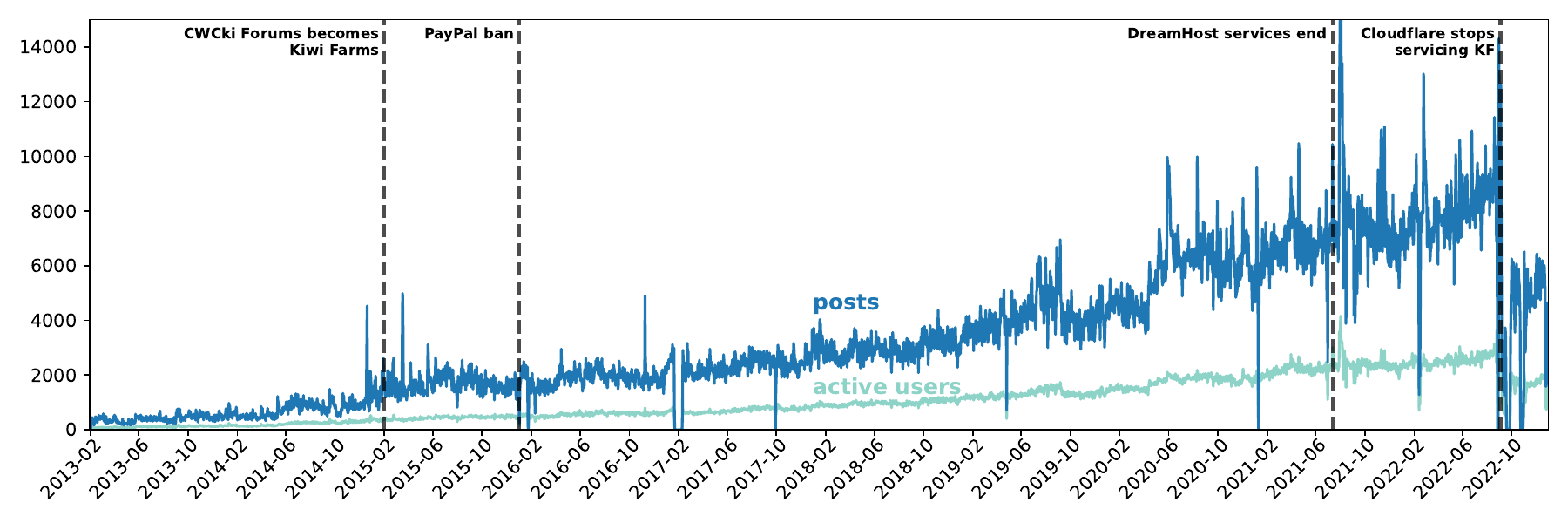}
    
    \caption{Daily number of posts and active users from the creation of Kiwi Farms (originally as CWCki Forums) until the end of 2022.}
    \label{fig:timeline}
\end{figure}

\subsection{Case study}
\label{sec:case-study}
Extremism and hate are known to fester and propagate on niche internet message boards. These have been the source of some of the most well-known harassment campaigns such as Gamergate in 2014. They have also played a critical role in bringing extreme violent right-wing ideology closer to the mainstream \cite{hineKekCucksGod2017}. While most of this type of activity takes place within individual channels of broader message board platforms, as has been the case with 4chan and, at one point, Reddit, it is also sometimes the case that a platform pops up and hosts almost exclusively this type of vitriol. This is the case for Kiwi Farms, which initially branched out from a combination of 4chan's \texttt{/v/} board, Encyclopedia Dramatica, and the Something Awful forum in 2008. This offshoot of these web spaces was a wiki page called ``the CWCki'', to collect information for trolling and harassing fandom artist Christine Chandler (who also goes by Chris-Chan) \cite{plessKiwiFarmsWebs2016}. The CWCki became insufficient for the desires of some users, whose communication on the CWCki's talk pages extended beyond the intended collaborative editing use-cases of a wiki talk page, prompting a former 8chan administrator known as ``Null'' to create the message board ``the CWCki Forums''. 

\begin{figure}[h]
    \centering 
    \includegraphics[height=.3\textheight]{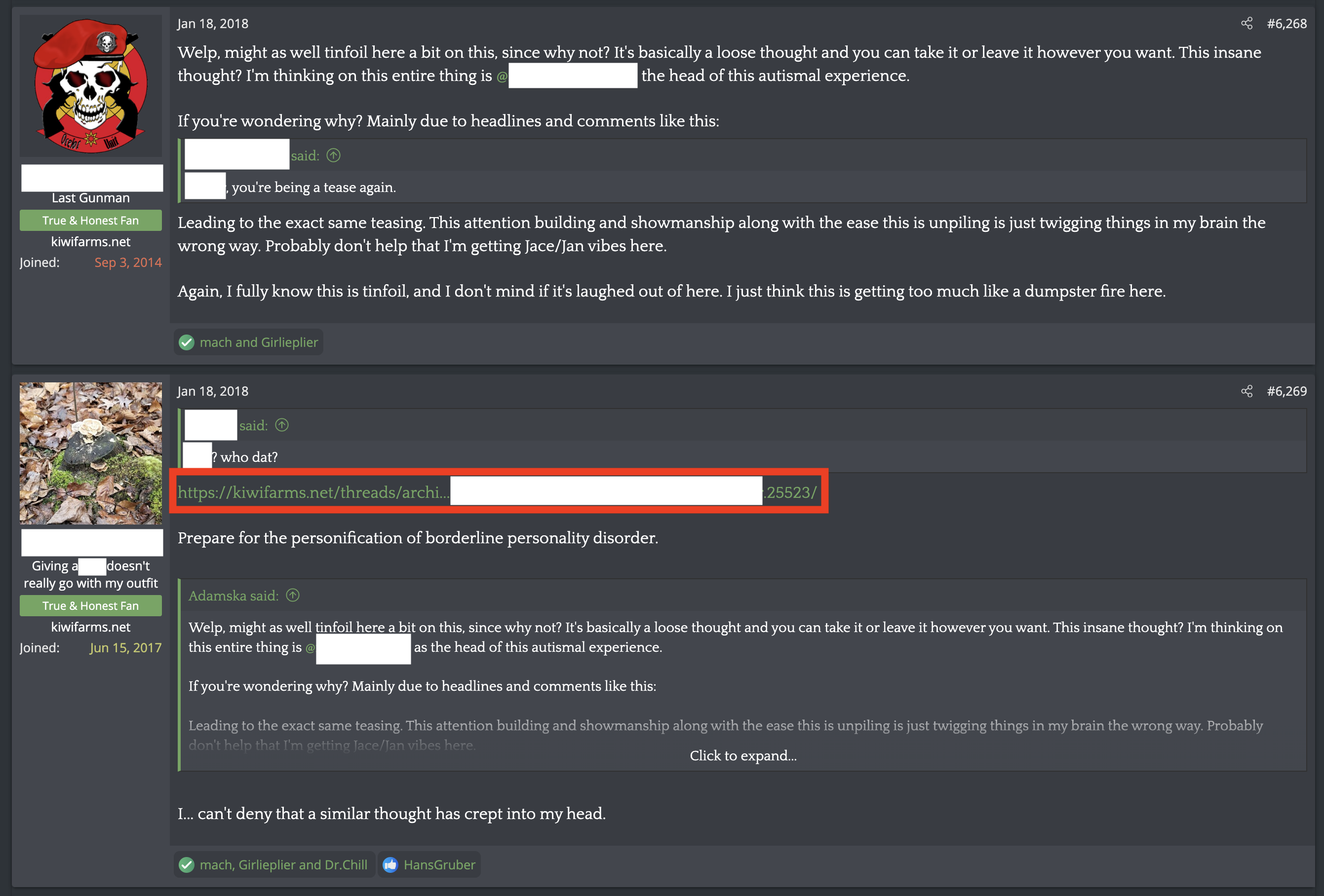}
    
    \caption{A screenshot of a few posts in a Kiwi Farms thread. A link to another thread is marked in a red rectangle. There are also quotes of previous posts and tagged mentions of other users, using the \texttt{@} character. Names, handles, and profanity are obscured with white boxes.}
    \label{fig:platform-constructs}
\end{figure}

Kiwi Farms became the official name of the platform as an alternative pronunciation of ``CWCki forums'', as the platform expanded to discussing and harassing people other than Chris-Chan, especially individuals with online presences that can be criticized through various bigoted logic, especially ableism \cite{vanschenckReplatformizationExpansionAlttech2026}. Users of the platform began referring to such targets as ``lolcow''s. Some examples of the way the community uses this word include
\begin{displayquote}
She is a grown a*s 30 year old woman who chose to move in with the most prolific lolcow in internet history. She knew exactly what she was getting into.

$-$ Kiwi Farms user, 2019
\end{displayquote}
and
\begin{displayquote}
**** will never have functional relationships with any of her family, nor is she capable of making and retaining friends in anything but the shortest of terms. **** wants everything to always be 100\% her way, and is completely unwilling to make any effort to change. She's one of the angriest and most actively unpleasant lolcows.

$-$ Kiwi Farms user, 2020
\end{displayquote}

The Wikipedia page for Kiwi Farms lists eight harassment campaigns of different targets, three of which have lesd to the deaths by suicide of the harassment victim. Strategies deployed by users of Kiwi Farms include the publication of private information about the target (doxing), sending false reports of imminent danger to police to have them deployed at the location of the target (swatting), and the mass-reporting of their social media accounts to shut down channels through which the victim would get support.

The growth of Kiwi Farms brought repeated pressure from service providers, including a PayPal ban in 2016 and the loss of XenForo and DreamHost services in 2021. A major \#DropKiwiFarms campaign followed the 2022 harassment of transgender streamer Keffals: Cloudflare stopped servicing the site on September 3, 2022, triggering further blacklisting, cyberattacks, data leaks, and substantial declines in users and activity. Despite these efforts, Kiwi Farms survived through dedicated infrastructure, cryptocurrency funding, bootlegged software. This process, dubbed ``replatformization'' by Van Schenck, partially constitutes the collective identity of Kiwi Farms to date \cite{vanschenckReplatformizationExpansionAlttech2026}. We include daily activity measurements with some notable changes in the platform in Figure \ref{fig:timeline}.

\subsubsection{Platform constructs}\label{sec:platform-constructs}

Kiwi Farms follows the standard formatting of a forum or message board. Users who create a new thread must choose the board where the thread will be situated. Boards are organized into categories, so a thread's general topic can be understood by its full directory, following the format \texttt{/Home/\{category\}/\{board\}/.../\{thread\}}. Some boards have further subdirectories where topic disambiguation is found to be necessary, but each lowest-level directory is given a unique board ID, and each thread has a unique ID as well. Threads are made up of individual posts, and these posts may include internal or external links, as well as quotes of previous posts in the thread. We show some of these post features in Figure \ref{fig:platform-constructs}.

\begin{figure}[h]
    \centering 
    \includegraphics[height=.3\textheight]{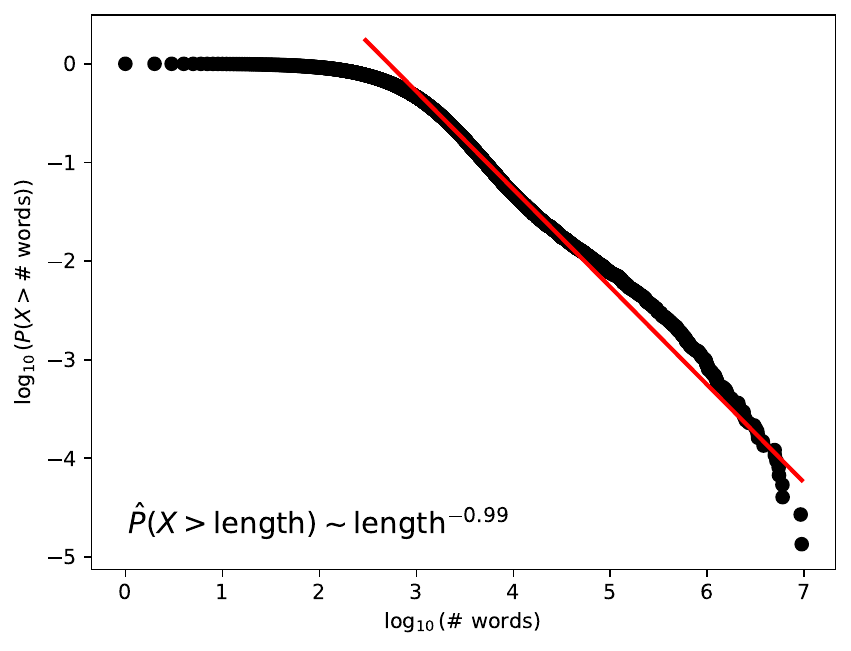}
    
    \caption{The complementary cumulative distribution function of document length for the threads contained within the dataset of Kiwi Farms posts in 2018 and 2019. The red line shows the scaling exponent $\gamma$ across values beyond an estimated lower limit of scaling, $x_{\text{min}}$. Both $\gamma$ and $x_{\text{min}}$ are determined by maximum likelihood estimation \cite{alstottPowerlawPythonPackage2014,clausetPowerLawDistributionsEmpirical2009}. Fit comparisons for other model distributions are shown in Supplementary Table \ref{tab:dist-comparisons}.}
    \label{fig:ccdf-doclength}
\end{figure}

\subsection{Data}\label{sec:data}
The entire historical archive of Kiwi Farms is included in the ExtremeBB dataset, collected by the Cybercrime Center at Cambridge University
\cite{vuExtremeBBDatabaseLargeScale2023}. Posts are collected as rows in a PostgreSQL database, which include the textual content of the post, the date, and unique identifiers for the post, creator, and thread. For thread-level analysis, we collapse posts with the same thread ID into single long-form strings. These cumulative threads have a mean length of 9475 words with a standard deviation of 76804 words, with the longest thread containing almost four-million words. The distribution of word counts is heavy-tailed with a power-law scaling region between $10^3$ and $10^6$. The complementary cumulative distribution function has a scaling exponent of $-0.99$, as shown in Figure \ref{fig:ccdf-doclength}\footnote{A power-law with an empirically determined minimum scaling limit is the best fit of four heavy-tailed distributions we tested. Comparisons against the alternatives are given in Table \ref{tab:dist-comparisons}.}.

\begin{figure}[h]
    
    \centering 
    $\vcenter{\hbox{\includegraphics[width=.5\textwidth]{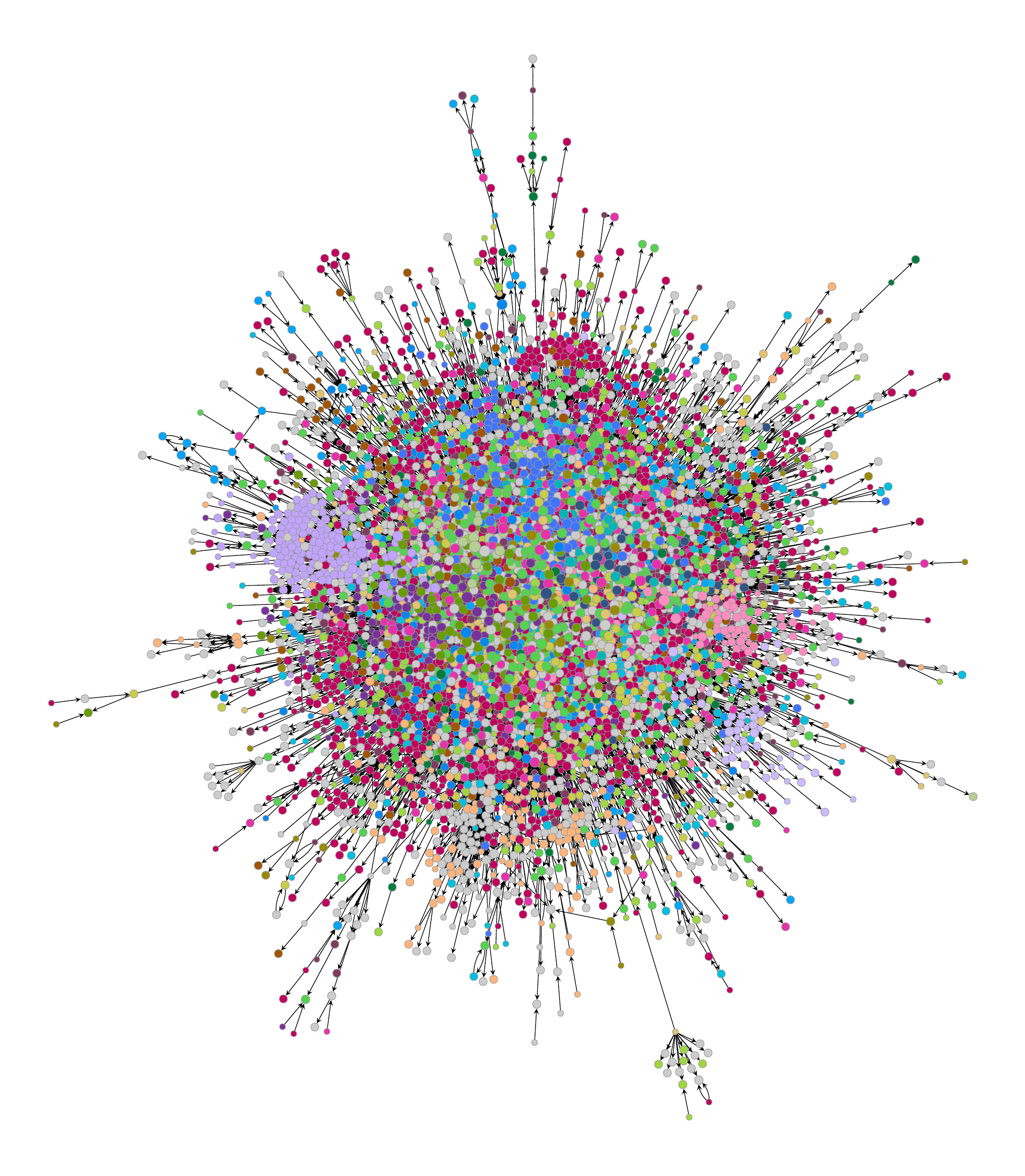}}}$
    $\vcenter{\hbox{\includegraphics[width=.45\textwidth]{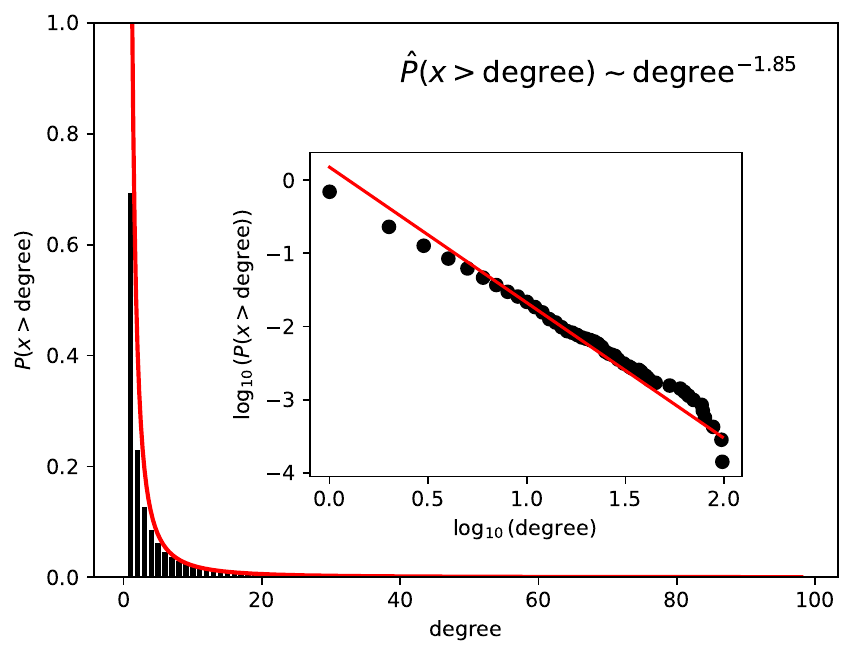}}}$
    
    \caption{(Left) The internal link network of Kiwi Farms from 2018 through 2022. Nodes are threads and edges are hyperlinks posted in one thread leading to another thread. Nodes are sized according to in-degree and colored categorically by their board ID. We describe some example boards in Section \ref{sec:temp-link-network} and include the full legend of corresponding board IDs in Supplementary Figure \ref{fig:network-legend}. (Right) The complementary cumulative distribution function of node in-degree in linear space. The model distribution indicated by the red line is determined with maximum likelihood estimation. We also include an inset showing the same data and fit in log-log space.}
    \label{fig:internal-network}
\end{figure}

\section{Results}\label{sec:results}

\subsection{Thread link network}\label{sec:temp-link-network}

To address RQ1 we operationalize the complex connectivity of the platform by constructing the network of threads on the message board connected by hyperlinks posted in threads that lead to others. This network is composed of the internal links and its degree distribution are shown in Figure \ref{fig:internal-network}. The nodes in this network are threads, and edges are hyperlinks between threads. Users may link to either a thread or a post in a thread, and in both cases it is considered a link to the thread. The nodes are colored according to the board in which they reside. 

\begin{figure}[h]
    \centering 
    \includegraphics[width=.9\textwidth]{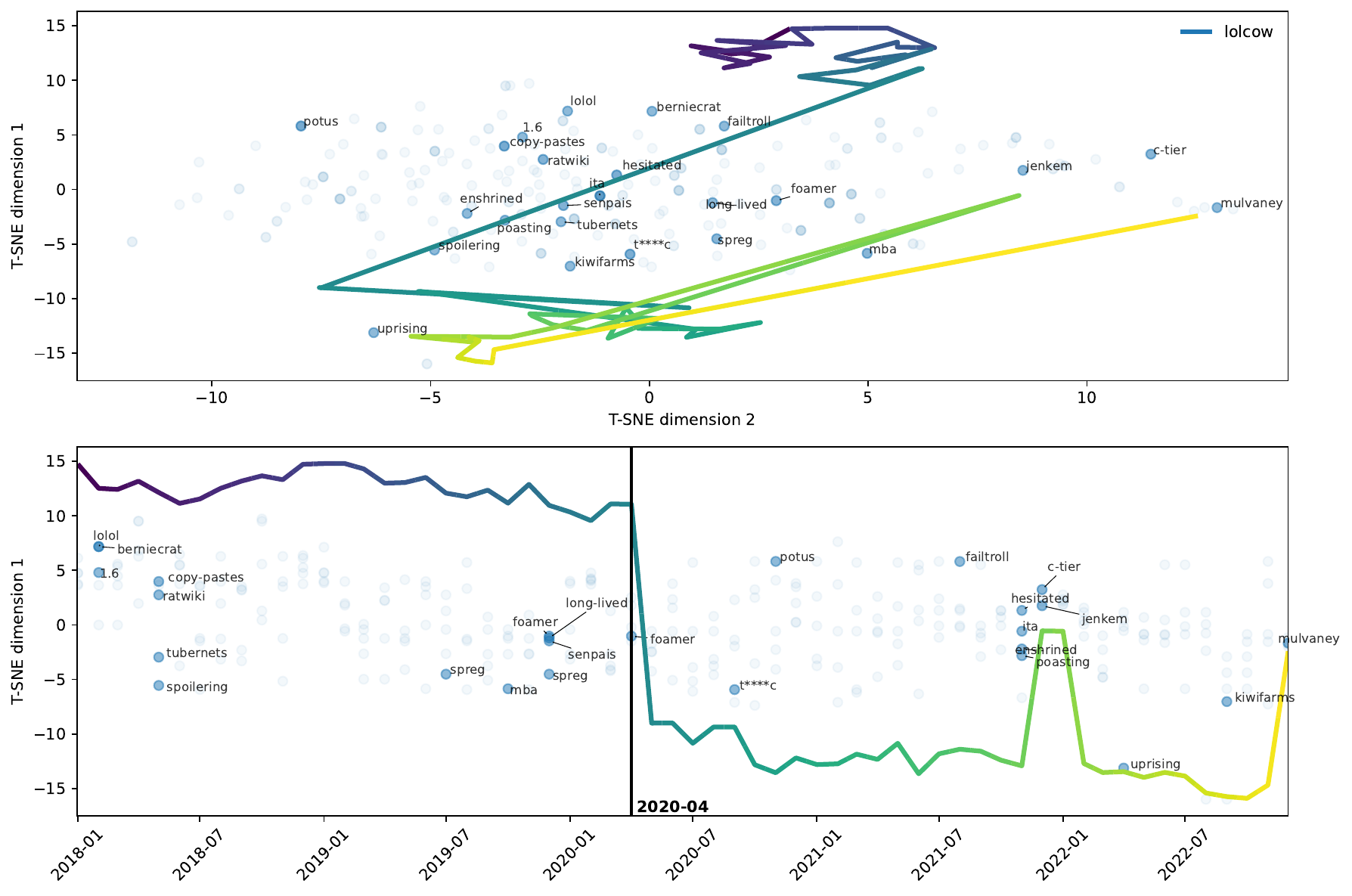}
   
    \caption{(Top) An example Gensim model fit showing the position of ``lolcow'' every month in two T-SNE components of the word2vec embedding space. A subset of the most similar words is labeled. (Bottom) The second T-SNE component plotted with time on the horizontal axis. The censored word is a slang term for people with autism that can be considered offensive in some contexts.}
    \label{fig:temp-embedding}
\end{figure}


We observe that the degree distribution of this network has a heavy tail, and report the power-law scaling exponent, with the caveat that this distribution does not meet the typical assumptions for such a fit, namely that the range of observed values only covers two degrees of freedom. With such a low maximum observation and a divergence from the power law fit at those higher values, this network is unlikely to be best described as scale-free, as true examples of scale-free networks are rare in the social and natural world \cite{broidoScalefreeNetworksAre2019}. In this case, however, one could postulate a preferential attachment mechanism for linking to internal threads on Kiwi Farms, as links to a thread are public and viewed by people who themselves are capable of creating a new link themselves. Such a mechanism is complicated by a variety of interacting sociotechnical factors, some of which are observable while others are not. The global assortativity coefficient for board ID is 0.45, indicating a greater-than-chance likelihood that any given link connects threads from the same board \cite{newmanMixingPatternsNetworks2003}. The most notable feature of this assortativity, however, is its heterogeneity throughout the network. We notice especially strong assortative mixing in certain regions of the largest connected component (LCC). The dense cluster of lavender (\patch{lavender}) nodes on the outer left of the network, and the slightly more diffuse groups of peach (\patch{peach}) and dark purple (\patch{darkpurple}) nodes consist of threads on the ``Prime Cuts'' board, which houses multiple threads centering on three people, whose online presence has drawn too much attention to be contained to a single thread. The dark blue (\patch{darkblue}) nodes that have a presence near the center are threads targeting trans people. Other boards exhibit disassortativity, such as the ``lolcows'' board, whose threads are shown as lime green (\patch{limegreen}) nodes that are scattered around the network.

\subsection{Temporal word embedding}\label{sec:results-temp-embedding}

We approach RQ2 by fitting temporal semantic models of platform-specific language. In particular, because this community consistently uses a nonstandard word, ``lolcow'', to describe their targets, changes in meaning of this word reflect changes in the way the community frames its targets, and makes sense of their actions against them. We show the monthly position of the word ``lolcow'' in an example Gensim embedding space decomposed into one and two t-distributed stochastic neighbor embedding (T-SNE) dimensions in Figure \ref{fig:temp-embedding}. We observe that this embedding position traverses the space of words, with a large relative jump from March to April 2020. It is important to note that the points represent the most similar words in the full dimensionality space of the word embedding. Because the plots only show one and two dimensions of the embedding space, the set of points that are similar to ``lolcow'' on that day may not be proximal to one another in the reduced space. 

\subsubsection{Ousiometric analysis of similar words}\label{sec:results-temp-ousiometry}

\begin{figure}[h]
    \centering 
    \includegraphics[width=.9\textwidth]{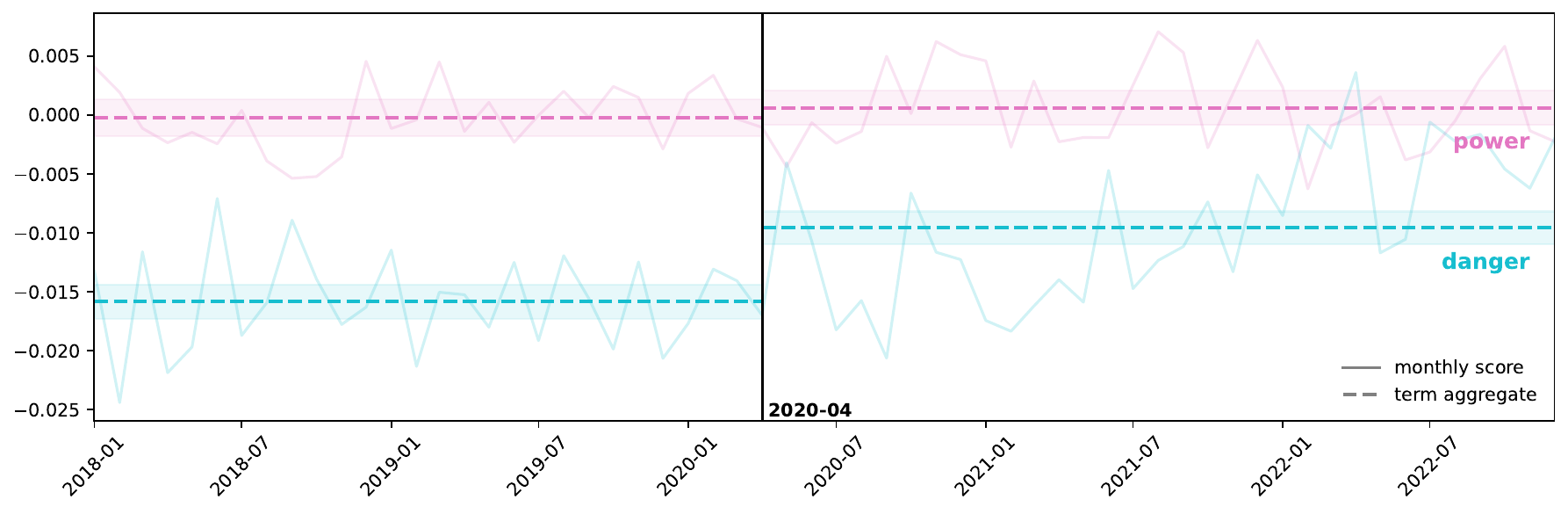}
    
    \caption{The mean ousiometric power and danger scores of the $1000$ most similar words to ``lolcow'' each month in the corpus. We include the aggregate means of two time periods within this larger window: before and after the onset of COVID-19 safety restriction in the United States in April 2020. These means and their 95\% confidence intervals are shown with the dotted lines and shaded regions.}
    \label{fig:temp-ousiometry}
\end{figure}

\begin{figure}[h]
    \centering 
    \includegraphics[width=\textwidth]{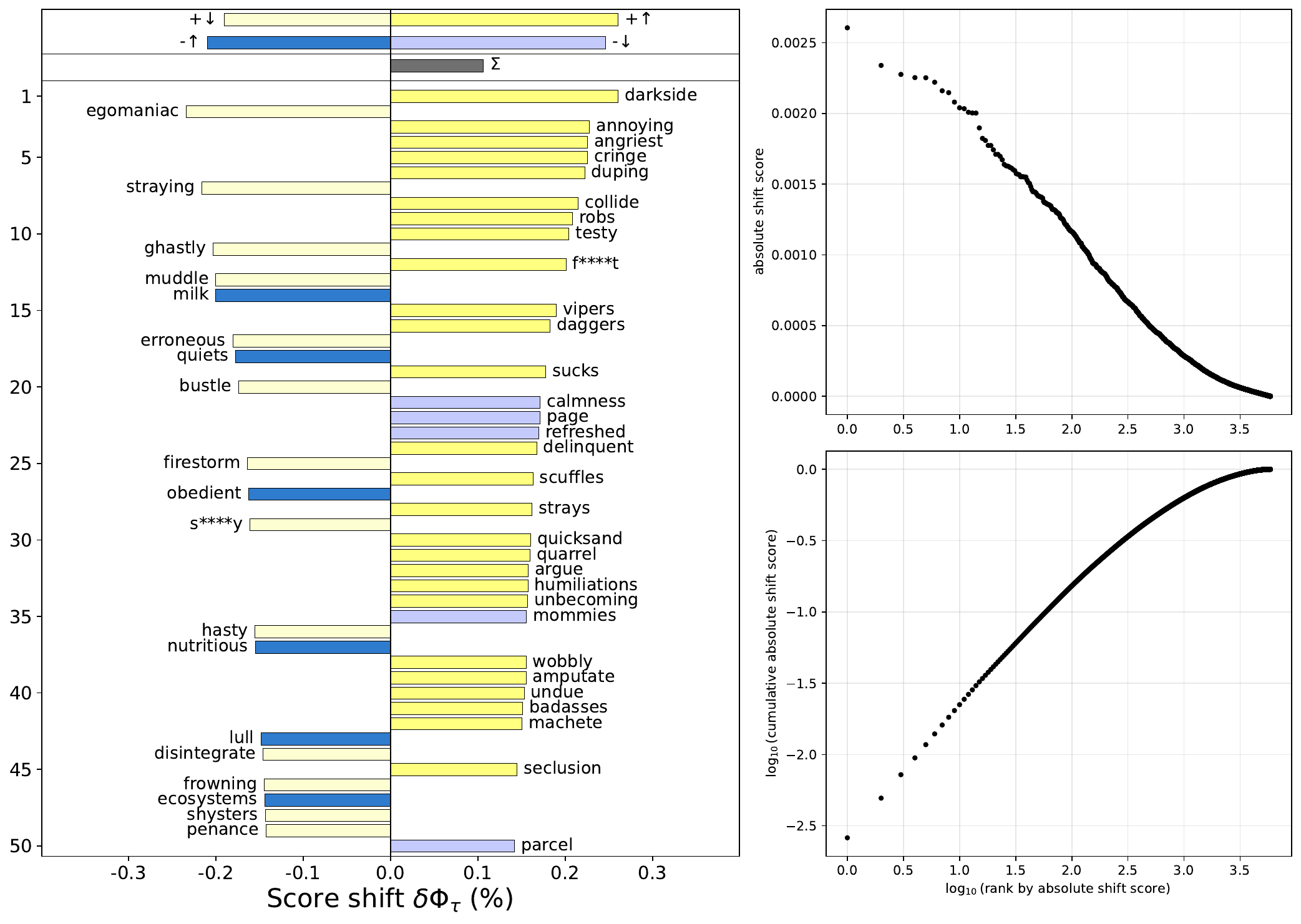}
    
    \caption{(Left) Ousiometric word shift: a bar chart showing the words that contribute most to the increase in danger of the aligned words after April 2020. Bars extending to the right indicate words making the context beginning on 2020-04-01 more dangerous than that prior, either because they are dangerous words whose similarity increased (bright yellow), or they are safe words whose similarity decreased (muted blue). Bars extending to the left are those contributing to making the context after 2020-04-01 safe, consisting of safe words whose similarity increased (dark blue) and dangerous words whose similarity decreased (muted yellow). The top bars indicate the relative sums of bar heights for each type, and the resulting net, marked with a $\sum$. Slurs and profanity are censored. (Top right) magnitude of contribution to the shift of each word plotted against the logarithm of the word's rank by magnitude. (Bottom right) Logarithm of the cumulative contributions plotted against the logarithm of rank.}
    \label{fig:danger-shift}
\end{figure}

The average Ousiometric scores of the similar words to ``lolcow'' reveal changes in the meaning conveyed by the word over time. We present in Figure \ref{fig:temp-ousiometry} the average danger-similarity (product of the danger of a word and its similarity to ``lolcow'' in the embedding space, with scores sourced from Dodds et al. \cite{doddsOusiometricsEssenceMeaning2026}) and power-similarity of the $1000$ most similar words, the computation of which is described in Section \ref{sec:methods-ousiometry}. While power occupies roughly the same dynamic range over time ($1.4\ \times$ standard error), we see that danger increases ($11.2\ \times$ standard error). This implies that the essential meaning of ``lolcow'' as used by this community has become more dangerous. While we choose to highlight April 2020 when describing the increase in danger-similarity for the date's exogenous importance as the start of COVID safety restrictions in the United States, we acknowledge that the measure continues to increase across time. We show this increase as a function of the chosen break-point date, as well as that for power-similarity, in Supplementary Figure \ref{fig:sliding-breakpoint}. It is especially notable that the danger-similarity increases persistently in late 2021 through 2022, aligning with successful campaigns by targets of Kiwi Farms to have service providers stop working with the platform. We further discuss the potential implications and causes of this in Section \ref{sec:discussion}. We break down this change in danger into its contributions from the 50 most influential words on this difference in the word shift plot shown in Figure \ref{fig:danger-shift}. We observe that the decay in the magnitude of word contributions to this difference is slow: the first 100 words contribute about 18\% of the total absolute divergence, with the first 1000 words contributing about 56\%. The rest of the divergence is accounted for in the remaining 2042 words. This indicates that this difference is driven a broad change in the vocabulary used, not simply a few words becoming more popular to use.

\subsection{Cohort binning}\label{sec:cohorts}

Changes in language surrounding the framing of the target provoke questions about the underlying mechanisms that produce them. In particular, one wonders whether the changes are a result of changing behavior of an existing user-base, or introduction of new, differently behaving users to the platform. To understand further what accounts for these longitudinal shifts, we examine the posting volume, reply networks, and ousiometric trends at the cohort level.

\begin{figure}[h]
    \centering 
    \includegraphics[width=.9\textwidth]{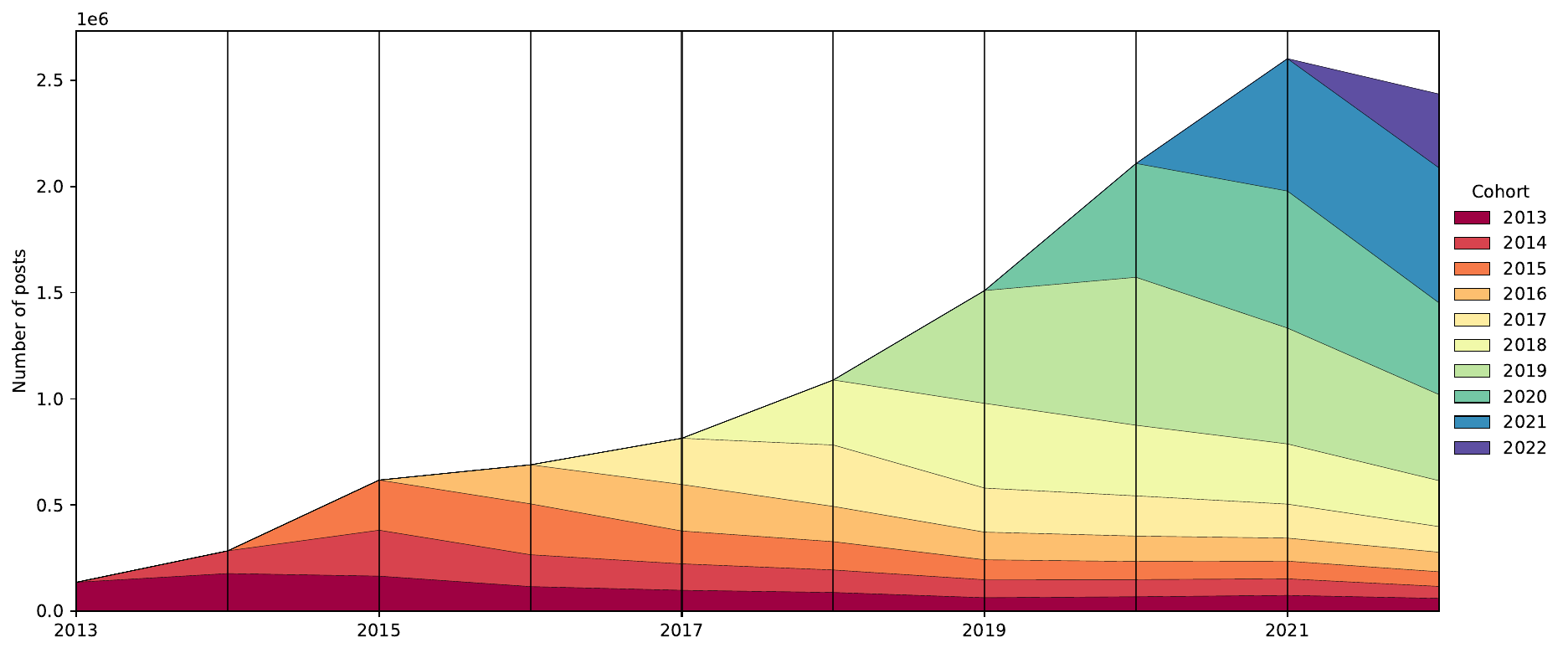}
    
    \caption{Area plot describing the volume of posts published by users of each cohort during each year from 2013 through 2022.}
    \label{fig:cohort-area}
\end{figure}

The cohort assigned to each year of the existence of Kiwi Farms is defined as the set of users who published their first post on the platform during that year. Figure \ref{fig:cohort-area} shows the volume of posts authored by users within each cohort on each year, beginning with the year of its creation in 2013. We extend the study period back to the beginning of the platform in 2013 to avoid technical issues that arise when cohort analysis uses a truncated dataset of platform activity\footnote{Using truncated data for cohort analysis leads to systematically erroneous cohort identification, as the first post by any given user could be in the period before study. This leads to an extreme inflation of the first cohort in the truncated study period, and smaller inflations to those later. By expanding to the full history of the platform, we ensure the accuracy of the first post, although uncertainty remains as to the amount of time a user spent reading the platform and interacting with it before their first post.}. Every cohort exhibits some persistence, most decaying to no less than 50\% of their original volume. We note a rise in the rate of increasing volume beginning in 2018, mostly due to new cohorts authoring more posts than those previously. This increasing rate continues until the transition from 2020 to 2021, at which point it slows down and then decreases in 2022 for the first time in the life of the platform.

\begin{figure}[h]
    
    \centering 
    
    \includegraphics[height=.27\textheight]{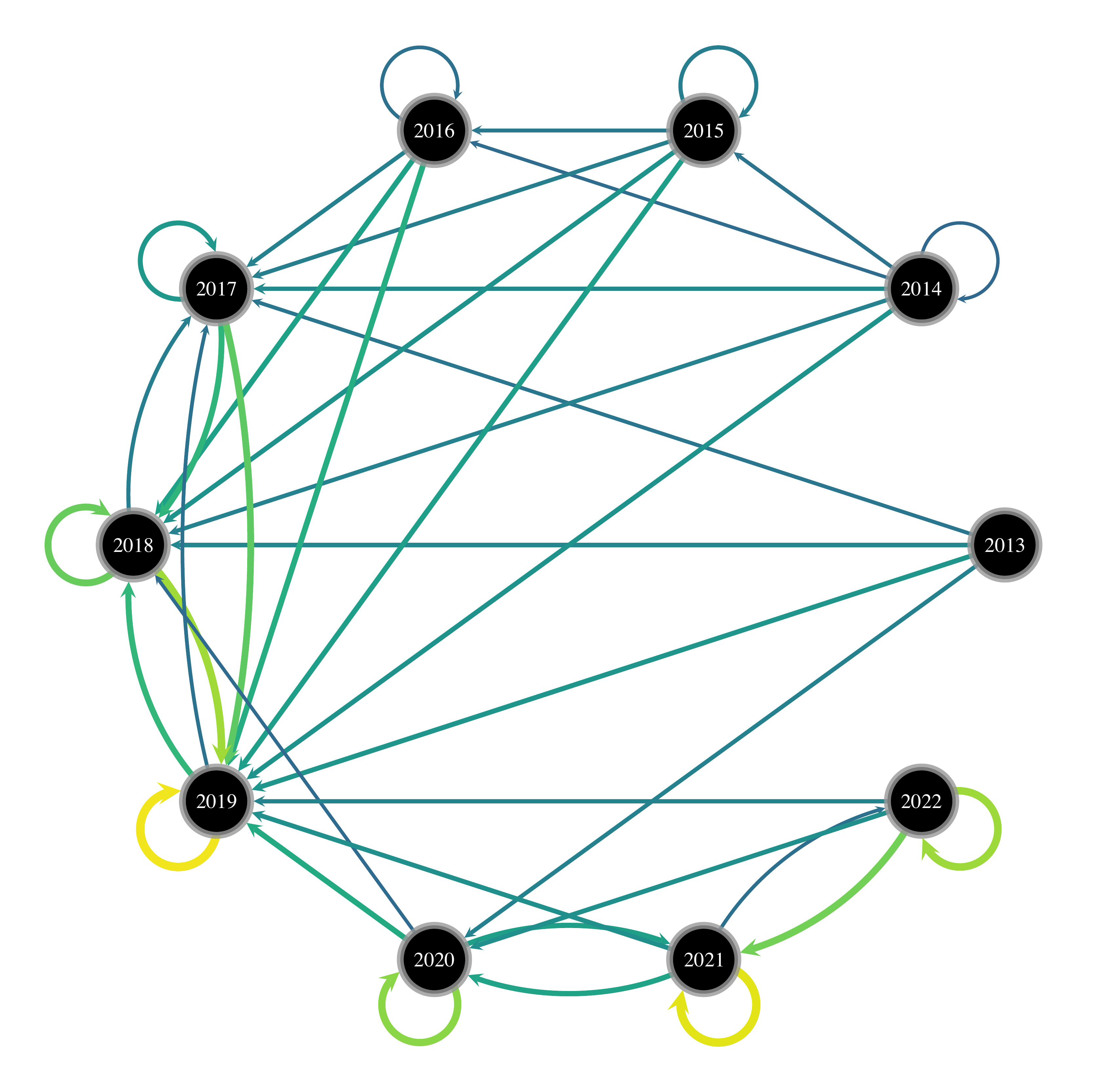}
    \includegraphics[height=.27\textheight]{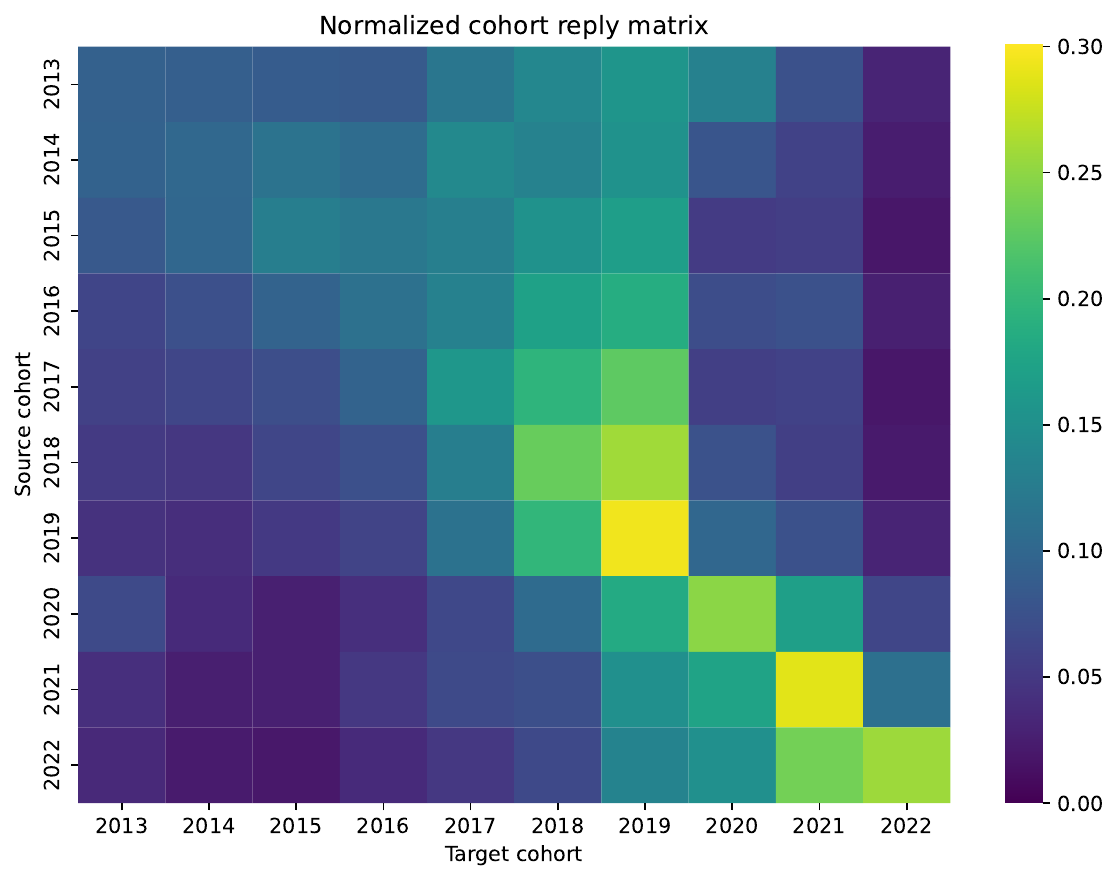}
    
    \caption{Cohort interaction rates, normalized by source cohort (The out-links sum to 1 for each source node, the sum of values for each row of the matrix equivalently sum to 1).}
    \label{fig:cohort-replies}
\end{figure}

We examine the level of interaction between cohorts by counting replies, displayed in Figure \ref{fig:cohort-replies}. Each link encodes the number of comments written by a user in the source cohort that reply to a user in the target cohort. We normalize the weights of these links by the total weight of all out-links originating from the cohort node, as early cohorts have more recorded activity within the nine-year span. We observe some adjacent-cohort reply activity as well as the expected pattern that early cohorts reply less to later ones, as there was less available time when those target cohorts existed within the lifetime of the source cohort. The reply matrix reveals a block pattern, where the later collection of cohorts beginning with 2020 reply mostly to each other, and cohorts 2013$-$2019 interact with one another.

\begin{figure}[h]
    \centering 
    \includegraphics[width=.9\textwidth]{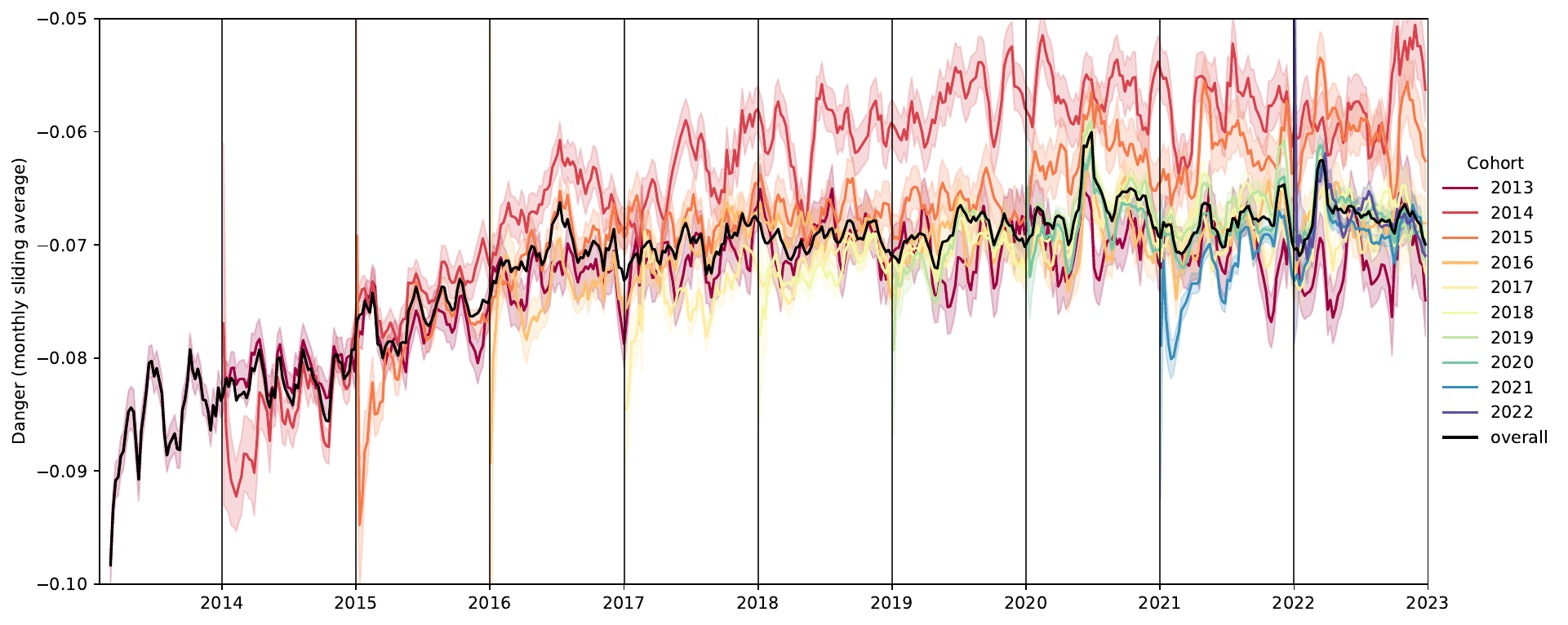}
    
    \caption{Temporal danger of words used by each cohort over the span of their time on the platform.}
    \label{fig:cohort-danger}
\end{figure}

To determine the extent to which the increase in danger found in Section \ref{sec:results-temp-embedding} may be explainable by new users, we examine the ousiometric danger of the lexicon used by each cohort over time. We show the ousiometric danger scores of the collections of posts written by each cohort within a 30-day time window sliding forward in increments of one week, accompanied by 95\% confidence bands in Figure \ref{fig:cohort-danger}.

We see that the overall trend in the danger of the whole text begins at its lowest relative point (-0.98), climbing over the first few years while approaching a plateau and remaining somewhat constant after 2018 (around -0.7). The danger of most cohorts appears to remain fairly close to the aggregate trend, not deviating much from one another. The major exceptions to this are the 2014 cohort, which reaches a significantly higher plateau than the aggregate, and the 2015 cohort, which exhibits a persistent jump occurring in early 2020. The trends reveal that any persistent rise in danger is not the result of a novel set of users who use language differently. The only persistent trend that could feasibly result in the increase in danger observed in Figure \ref{fig:temp-ousiometry} is that seen in the rise in danger of cohort 2015 starting at the beginning of 2020 (from a stable oscillation around -0.7 to a wider range of values jumping between -0.7 and -0.6).

\begin{figure}[h]
    \centering 
    \includegraphics[width=.9\textwidth]{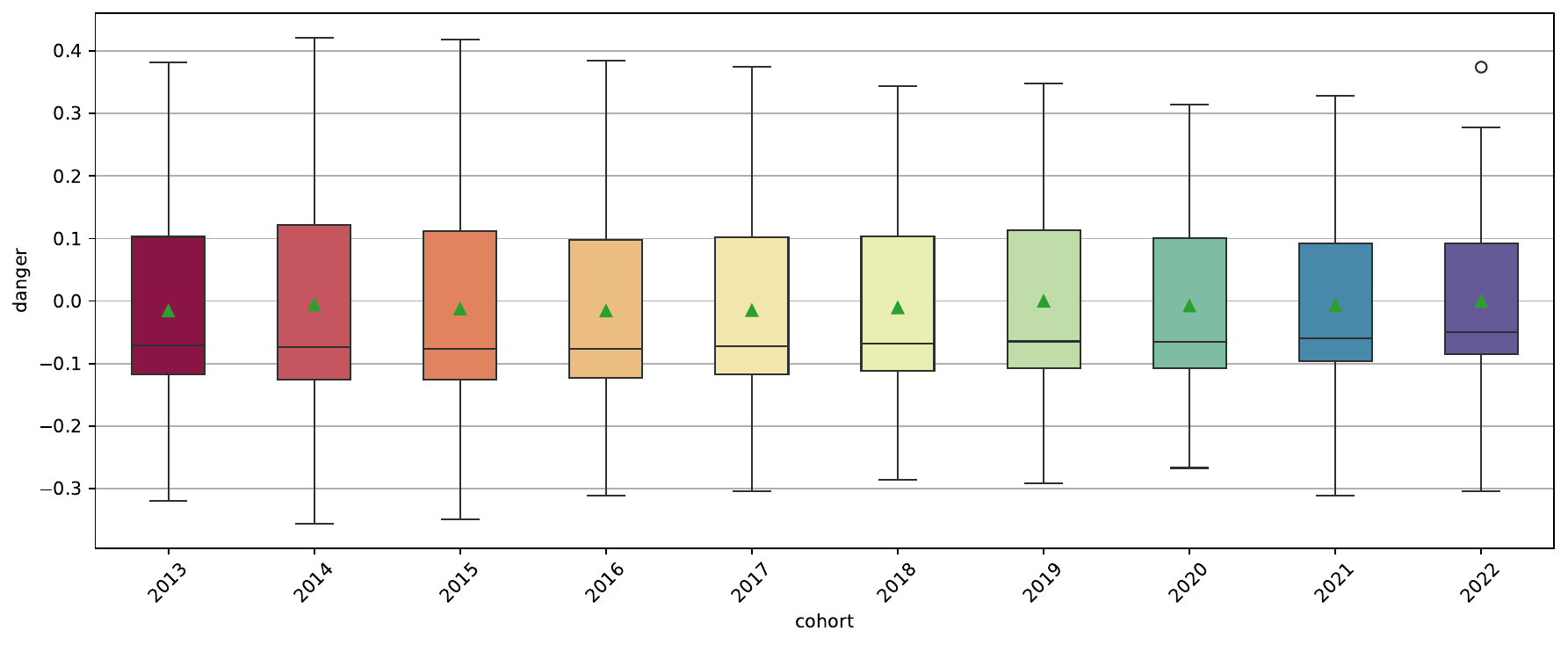}
    
    \caption{Danger-similarity distribution boxplots of most similar words to ``lolcow'' for each yearly cohort, over their tenures through 2022. Box edges and horizontal lines indicate quartiles, and triangles indicate means.}
    \label{fig:cohort-embed-danger}
\end{figure}

While we do not observe a strong signal that a single cohort's overall vocabulary drives the change observed in \ref{sec:results-temp-embedding}, the possibility that an individual cohort discusses harassment targets differently remains. We thus examine the results of a refit word2vec model with the word ``lolcow'' annotated with the cohort year of its author, rather than the date of publication. We present the danger-similarity product distributions for the 1000 most similar words for each cohort in Figure \ref{fig:cohort-embed-danger}, which show no major differences between cohorts. This indicates a change in framing given by existing users, rather than the introduction of new users who write about harassment targets differently. We include a similar plot cohorting by month implying the same conclusion in Supplementary Figure \ref{fig:embed-monthly-cohorts}.

\subsection{Thread decay}

\begin{figure}[h]
    \centering 
    \includegraphics[width=.3\textwidth]{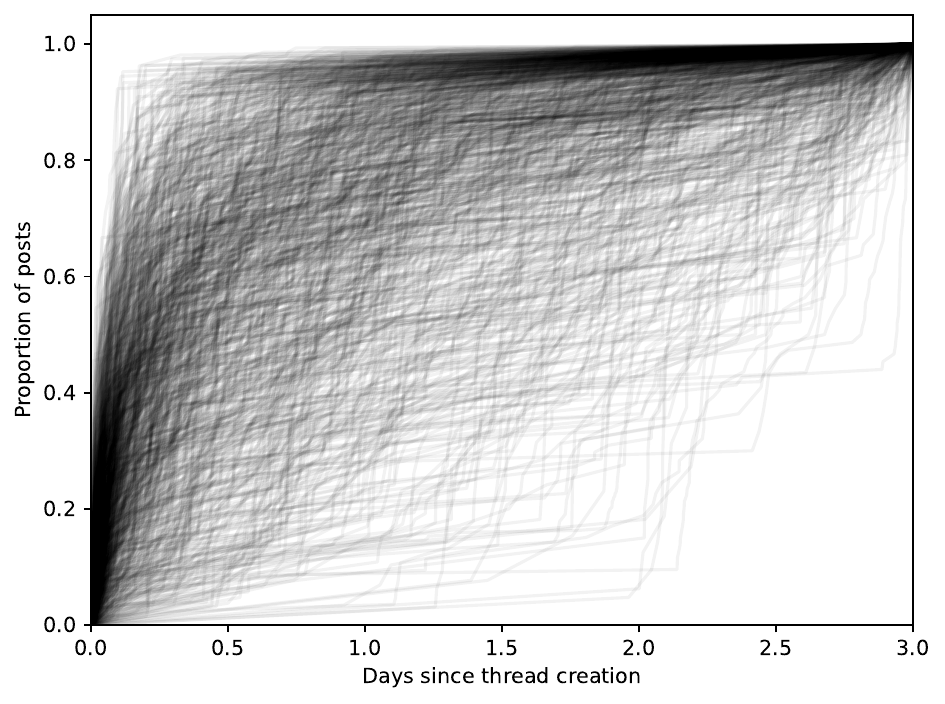}
    \includegraphics[width=.3\textwidth]{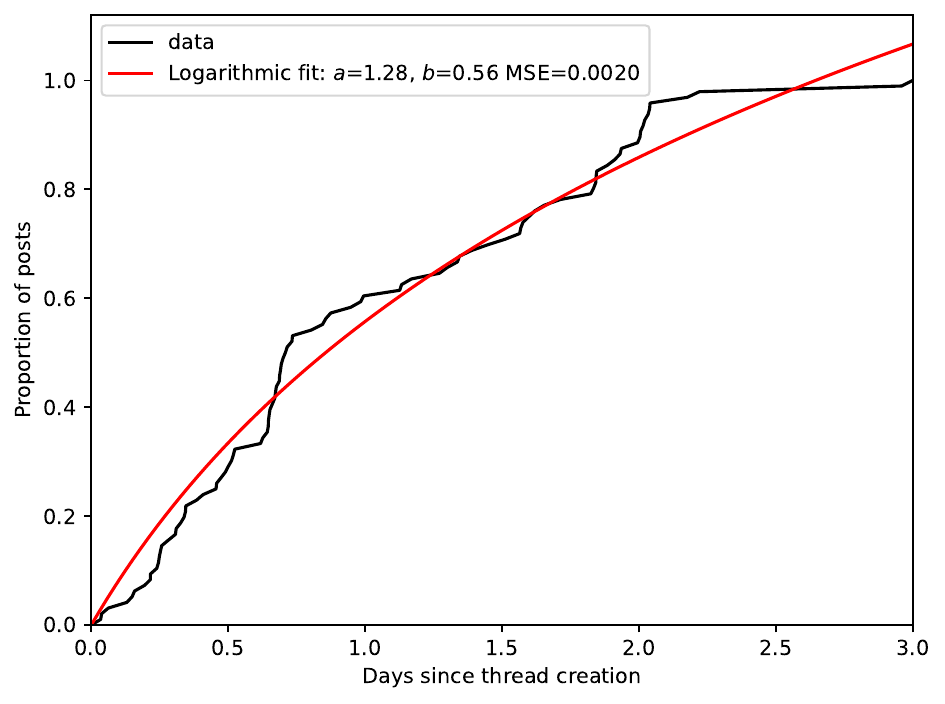}
    \includegraphics[width=.3\textwidth]{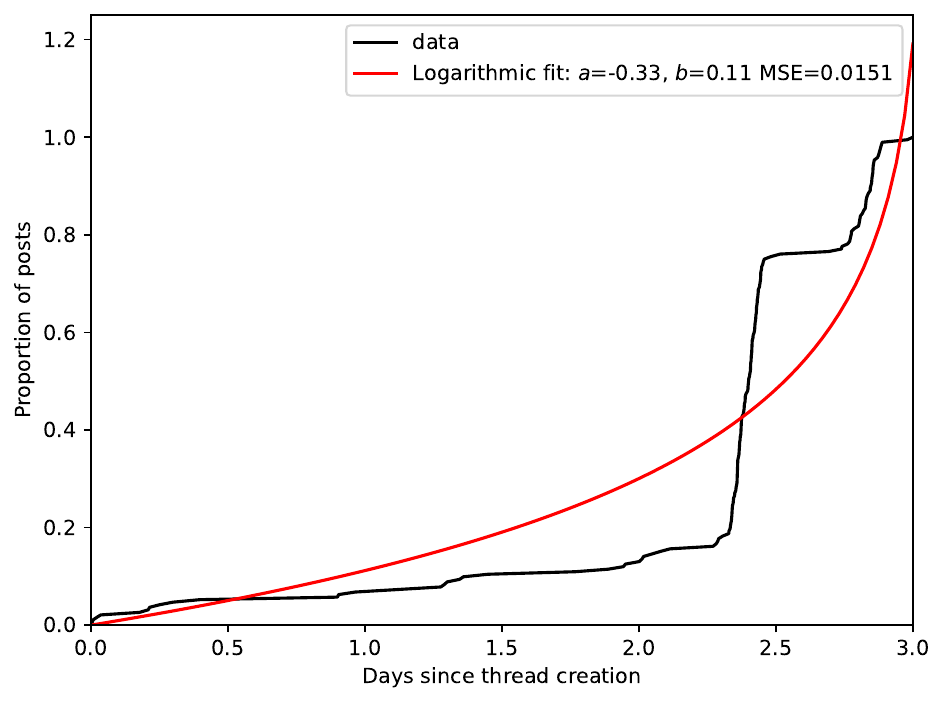}
    
    \caption{Thread growth traces and model fits. (Left) A random sample of 1000 traces, normalized to reach 1.0 at the end of the third day. (Center) A sample trace fit to the logarithmic growth model with a positive $a$. (Right) A sample trace fit to the model with a negative $a$ value.}
    \label{fig:traces-example-fits}
\end{figure}

We address RQ3 by fitting early growth behavior of threads on Kiwi Farms to nonlinear models, and examining the relationship between the model parameter and long-term thread size. Figure \ref{fig:traces-example-fits} shows the process of analyzing the cumulative post counts in threads over the course of the first 3 days since their first post. Each trace is standardized to be 1 at 3 days. We show a sample of 1000 of these traces and two examples of traces fit to the logarithmic curve
$$T^\ell_{a,b}(t) = b\cdot\frac{\log(a\cdot t+1)}{\log{(a+1)}}.$$
Lower values of $a$ correspond to a slower decay of the growth rate. When $a<0$, the growth rate increases over time during that three-day window \footnote{We acknowledge that in the $a<0$ case, this is not a physical growth model as there is a finite limit to $t$ beyond which the function becomes undefined. We, however, keep the model as it is not intended to forecast growth, and the value of $a$ is useful for telling us that the growth increases over time in these cases.}. We compare this parameter to the post count after 30 days. We show this relationship in Figure \ref{fig:thread-length-param}. We see that while the distribution at each of the logarithmically-spaced bins overlap significantly, these heavy-tailed distributions of 30-day thread size shift downward with higher values of the parameter $a$ in the logarithmic model fit. Notably, the average thread length for $a\leq0$ is 627 posts, whereas the mean for $10^2<a\leq10^4$ is 56. Because the distributions are heavy-tailed, the magnitude of these means is driven by relatively few very high values. The spread of these distributions also narrows as the value of $a$ rises. This implies that a low value of $a$ in the logarithmic growth fit in the early days of a thread indicates high potential for a viral thread, but the probability of this outcome remains low across all cases. 

\section{Discussion}\label{sec:discussion}

These analyses provide an empirical characterization of a complex interconnected system of actors whose interactions are dictated and mediated by a platform and whose affordances are one part intentionally designed, another part adopted from a lineage of internet message boards before it. Monitoring and understanding this platform is a matter of public safety, but reading the content of this platform in real time would be both resource intensive and deleterious to the mental health of the reader. We propose, however, that a more distant and semi-automated monitoring approach may balance these elements.

We note that several social and technical features of the system aid the very study of it. For instance the thread network can be easily constructed given that the thread identification number is present in the link to a thread or to any post within it. We are also able to monitor the shifting attention between targets of this community by leveraging their unique vocabulary. The fact that ``lolcow'' is a novel term used almost exclusively by this community to designate a target allows us to anchor on that word and track important movements semi-automatically. 

From qualitative analysis of the thread-link network, we observe that threads tend to be topologically situated by category. If a thread is about a specific person, it will be connected by links to threads about others with similar presentations of identity, and to discussion threads about the identity group to which the target belongs. We see this in regions such as a cluster of threads targeting members of the furry community and threads criticizing the community as a whole. We also note that the maximum in-degree of any particular thread in the network is in the area of 100 across time scales. A temporal subset of the network containing a single year's worth of data has nearly the same maximum in-degree as the full four-year network, indicating topical turbulence on the platform. Despite this, some threads still reach hundreds of thousands of posts, possibly meaning that at some point a largely static group of users becomes the sole set of contributors, who are nonetheless invested enough to continue the conversation beyond the point when new links to the thread are appearing elsewhere.

\begin{figure}[h]
    \centering 
    \includegraphics[width=.8\textwidth]{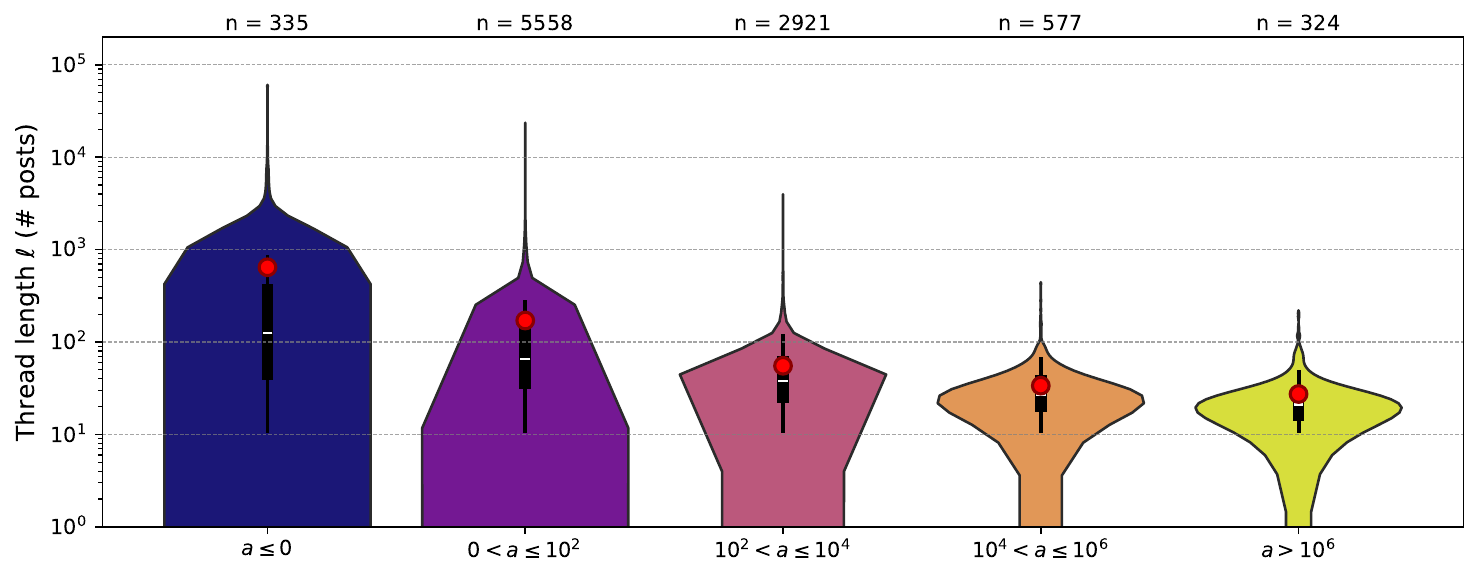}
   
    \caption{Violin plots showing the distribution of number of posts in each thread in the dataset after 30 days, split into logarithmically-spaced bins based on the value of $a$ for the best logarithmic growth fit. Box plots displaying the quartiles are shown within each violin, along with a red dot displaying the mean.}
    \label{fig:thread-length-param}
\end{figure}

Our temporal word embedding reveals that ``lolcow'' traverses a latent semantic meaning space over the studied time-window, but the shift in meaning accelerates over the course of 2020 and the years that follow. We find that the most proximal words indicate specific targets, but this jumping from target to target does not explain the longer directional movement across this lexical dimension. Our ousiometric analysis on this embedding space reveals a major component of this lexical change: an increase in the implied danger of a ``lolcow''. The jump to a consistent higher aggregate danger value of similar words reflects a broader trend across the public sociopolitical environment, where influential figures assert that those whose identities or activities fall outside the accepted mainstream pose a threat. Individuals or communities that would have once been considered merely strange or ``cringe'' are increasingly framed as menaces to society. The idea that this shift in rhetoric increases the likelihood of targeted violence toward the demonized group has become known as ``stochastic terrorism'' \cite{ammanStochasticTerrorismLinguistic2021,angoveStochasticTerrorismCritical2024}. In this way, the fringe message board Kiwi Farms adopts the same moral panic mechanism as more mainstream political movements and media.

It is also notable that this rise in invoked danger coincides with the victims of this mob harassment fighting back, and securing material setbacks for Kiwi Farms as a platform. Most notably, Keffals' \#DropKiwiFarms campaign resulted in significant drops in the hosting ability of the platform over the course of 2022. This along with the broader changes in society brought by the COVID-19 pandemic are potential factors contributing to these changes.

Seeing the essential meaning of a unique word used to describe a community's targets become more dangerous is troubling, and provokes one to question whether this reflects a change in perspective of the existing members of the community or the entrance of new members to the community who feel differently. Our cohort analysis suggests that the mechanism is primarily the former, as individual new cohorts do not appear to use more dangerous language. 

With the prior analyses describing the mechanisms of this distributed sociotechnical system and how it is situated in the broader sociotechnical environment, we are equipped to contextualize the final analysis: our modeling of temporal thread trajectories. As is the case across human and non-human complex systems, the distribution of a critical output, in this case thread-length after 30 days, is extremely heavy-tailed. Highly viral threads are thus extremely rare, and no single early signature is a strong predictor of long-term performance. As we have shown, however, it is possible to classify the risk that a thread is going to go viral by the growth function of best fit in the first three days. Figure \ref{fig:thread-length-param} shows that threads whose $a$ value in the growth function is negative have an average 30-day thread length that is over 600, while those with a value between 0 and 100 have an average 30-day length around 100, with the average thread length only getting shorter as the parameter value increases. In practical terms, any individual new thread targeting a person for online harassment is unlikely to go viral, but the early growth pattern allows us to classify the risk of extreme virality. Given the consequences of such virality, the value of such a threat assessment is clear.

\subsection{Ethical considerations}

When working with social media and message board data, there are several ethical issues that must be considered and navigated thoughtfully. While posts are public, internet users have a specific idea of the likely audience for what they say, and highlighting public posts in research exposes them to a different and potentially larger audience, possibly without the context that the original audience understands. 

This concern may seem misplaced, as the message board in question exists almost exclusively to broadcast personal information of vulnerable people. It is important to recognize, however, that there is a spectrum of involvement of individual users in the harmful activities for which Kiwi Farms is known. For this reason, we keep the majority of subjects anonymous, aggregate data into threads rather than posts, and remove the actual thread IDs and links in published analysis and supplementary data. We only mention individuals who are well known to have taken leading roles in harm and are documented in mainstream histories of the platform such as its article on Wikipedia. When taking this balanced approach, we refrain from exposing any novel individual activity while focusing on insights regarding the aggregate behavior that can inform monitoring and policy.

This approach also balances the safety of the research team with the need to investigate these questions and disseminate findings across the broader research community. We recognize that pursuing research on this subject can potentially put the researcher in the crosshairs of a community capable of and willing to impact the quality of life of its targets. For this reason, our research team has invested in personal cybersecurity measures when we publish this research.

\subsection{Limitations and future work}

It should be noted that the system in question can only ever be observed indirectly and incompletely. This is to say, collecting all public posts leaves out peer-to-peer messaging, private channels, in-person communication, and personal thoughts. All observations and subsequent theorizing are done with this in mind.

The other important acknowledgement is the mediation of interactions on the internet. This web of individuals spans many platforms, each with their own affordances and nuances. This mediation impacts the way individuals communicate. For instance, 4chan posts have no username attached to them, so compiling language and behaviors of specific individuals is impossible. Some users identify themselves, allowing some posts to be attached, but there is no way to know what other posts might be written by the same person, or whether they are being truthful about which historical posts they wrote. We do our best to uncover useful trends despite this mediation, but it is important to consider the limitation, especially when conducting work that includes multiple platforms. 

This study only considers components of the system within the platform, limiting the breadth of our lens. Kiwi Farms users are, like everybody, subject to influence from peer interactions as well as social and traditional media. Treating the platform as insular excludes important dynamics. In future work, we plan to include external links in the network, observing which platforms the largest threads contain links to.

\section{Conclusion}\label{sec:conclusion}

This study is grounded in a problem of broad societal importance: while targeted harassment from an online platform like Kiwi Farms directly affects relatively few people, it consumes the entire lives of those individuals and induces a chilling effect on the exercise of living as one's authentic self. In doing this work, we add to a body of knowledge that can inform policy or be implemented by watchdog groups in the case that the government does not prioritize the safety of vulnerable people. While we primarily aim to understand how these spaces operate at a systems level, this knowledge of digital traces that can identify an ongoing attack could prove useful to groups trying to mitigate these harassment campaigns in real time. In addition to policymakers and human rights groups, these insights and methods may be useful to social media safety teams. We know that some platforms have user-bases that are more susceptible to being maliciously exposed to hostile audiences on other platforms. For instance, the account LibsOfTikTok on X exists exclusively to provoke the anger of X users toward some users of TikTok. Likewise, many of the victims of Kiwi Farms harassment are targeted due to their activity on Tumblr. Knowing the patterns of behavior on Kiwi Farms and what signals to watch for could help safety teams more effectively defend the minoritized among their users.

\section{Methods}

\subsection{Internal link network}

It is common for these posts to include hyperlinks either to other websites or to other threads on Kiwi Farms. For internal links, the URL contains the unique identifier for the thread to which it links. For each post with an internal link, we extract the linked post identifier. Each hyperlink thus constitutes an edge from the thread it is posted in to the thread containing the linked post. For each edge, we preserve the date of the linking post, source thread, target thread, and the boards containing the two threads. We compute board assortativity as 
$$
\frac{\text{Tr}(\textbf{e}) - ||\textbf{e}^2||}{1 - ||\textbf{e}^2||}
$$
where \textbf{e} is the $n$x$n$ matrix, where $n$ is the number of edge types (unique boards), whose elements $e_{ij}$ are the fraction of all edges which connect a thread in board $i$ to a thread in board $j$ \cite{newmanMixingPatternsNetworks2003}.

We compute the decay exponent by binning the network nodes (threads) by in-degree using the \texttt{powerlaw} package in Python, and fitting the decaying curve with maximum likelihood estimation \cite{alstottPowerlawPythonPackage2014,clausetPowerLawDistributionsEmpirical2009}. We compute the layout using the default spring-force layout in \texttt{graph-tool}.

\subsection{Temporal word embedding}

We fit an ensemble of 50 Gensim word2vec word embedding models, with platform-specific slang decomposed into temporal components \cite{mikolovEfficientEstimationWord2013}. We do this in the preprocessing stage by concatenating each occurrence of the unigram ``lolcow'', slang for the targets of this community's harassment, with the year and month of the post containing the occurrence, such that all instances of the word occurring in January 2018 become ``lolcow-2018-01'', and so on. We then fit the models to this corpus with these labeled unigrams, noting their relative positions over time. Changes in this position over time indicate dynamic semantic associations with the term, potentially revealing shifts in the type of person being targeted, or movements to specific new targets. Our word2vec models are parameterized with $v \in \mathbb{R}^{100}$ and a window size of 100. We use this large window in order to capture topical and essential similarities more than syntactic relations. 

\subsubsection{Ousiometric analysis}
\label{sec:methods-ousiometry}

To further investigate these changes, we collect the most similar words to each month's instance of ``lolcow''. These are the words with the highest cosine similarity, $\cos \theta = x_1 \cdot x_2/|x_1||x_2|$. For each date, we take the words with the $1000$ highest average similarities for each month and compute the aggregate power and danger scores for each month using the ousiometric word score data set from Dodds et al. \cite{doddsOusiometricsEssenceMeaning2026}. To account for the potential instability of vector similarities across model initializations \cite{antoniakEvaluatingStabilityEmbeddingbased2018}, we use an intersection of similar words across 50 independently fit models. We pull these $1000$ words from the intersection of the sets of $100$-thousand most similar words on that date identified by each separately seeded model. To aggregate the power and danger score of the collection of words contextually aligning with ``lolcow'' in our model, we combine the word scores from the Ousiometric corpus with cosine similarity. The weighted score (power or danger) of the context in a particular month, $\Phi_t$ is then,
$$
\Phi_t = \sum_{w \in W_t} \phi_w \cos \theta_{w,t},
$$
where $\phi_w$ is the score of the word in the Ousiometric corpus, $W_t$ is the set of the 1000 most similar words by cosine similarity to ``lolcow'' at month $t$, and $\cos \theta_{w,t}$ is that cosine similarity for word $w$. To aggregate these monthly scores over a longer span of months $T$, we average this aggregate score across the months as
$$
\Phi^{(T)} = \frac{\sum_{t\in T}\Phi_t}{|T|} = \frac{\sum_{t\in T}\sum_{w \in W_t} \phi_w \cos \theta_{w,t}}{|T|}.
$$

With a procedure established Gallagher et al., we use these scores to compute word shifts by ordering words by their contribution to the aggregate difference in power or danger \cite{gallagherGeneralizedWordShift2021}. We do this by first determining the average similarity of the word in our model over the time-window. This measure, which functions analogously to a frequency, is computed as
$$
f_w^{(T)} = \frac{\sum_{t \in T}(\delta(w \in W_t)\cos \theta_w)}{|T|},
$$
where $\delta(w\in W_t)$ is a Kronecker delta function valued at 1 if $w$ is a member of $W_t$ and 0 otherwise. The contribution of a single word to the change in aggregate score between two time periods, $T_1$ and $T_2$, is then,
$$
\Delta\Phi_w = (f_w^{(T_2)}-f_w^{(T_1)})\phi_w.
$$

\subsection{Cohort binning}

We define each annual cohort as the subset of users whose first post on the platform is published within that year. To avoid erroneous assignments of users to later cohorts, we extend our studied period backward in time to the genesis of the platform. 

\subsection{Thread decay analysis}

Lastly, we examine the decaying growth rate for new threads on Kiwi Farms. We apply methods developed by Pfeffer et al. for analysis of tweets to analysis of these threads \cite{pfefferHalfLifeTweet2023}. We collect all of the threads posted to the ``lolcow'' board on Kiwi Farms within the studied period. We fit their cumulative number of replies over time for the first three days to the logarithmic curve
$$T^l_{a,b}(t) = b\cdot\frac{\log(a\cdot t+1)}{\log{(a+1)}}.$$
For each fit, we compute the mean-squared error, examining the relationship between the difference in these errors and the total number of posts to the thread after 30 days.
\backmatter

\bmhead{Acknowledgements}

B.F.E. appreciates conversations with colleagues at the Advanced Graduate Workshop on Computational Social Science at the Santa Fe Institute, especially those with Dakota Murray. 



\section*{Declarations}

The authors declare no conflict of interest.




\bigskip





\begin{appendices}

\bibliography{references}

@article{alstottPowerlawPythonPackage2014,
  title = {Powerlaw: {{A Python Package}} for {{Analysis}} of {{Heavy-Tailed Distributions}}},
  shorttitle = {Powerlaw},
  author = {Alstott, Jeff and Bullmore, Ed and Plenz, Dietmar},
  year = 2014,
  month = jan,
  journal = {PLOS ONE},
  volume = {9},
  number = {1},
  pages = {e85777},
  publisher = {Public Library of Science},
  issn = {1932-6203},
  doi = {10.1371/journal.pone.0085777},
  urldate = {2026-08-08},
  langid = {english}
}

@article{ammanStochasticTerrorismLinguistic2021,
  title = {Stochastic {{Terrorism}}: {{A Linguistic}} and {{Psychological Analysis}}},
  shorttitle = {Stochastic {{Terrorism}}},
  author = {Amman, Molly and Meloy, J. Reid},
  year = 2021,
  journal = {Perspectives on Terrorism},
  volume = {15},
  number = {5},
  eprint = {27073433},
  eprinttype = {jstor},
  pages = {2--13},
  publisher = {Terrorism Research Initiative},
  issn = {2334-3745},
  urldate = {2026-04-14}
}

@article{angoveStochasticTerrorismCritical2024,
  title = {Stochastic Terrorism: Critical Reflections on an Emerging Concept},
  shorttitle = {Stochastic Terrorism},
  author = {Angove, James},
  year = 2024,
  month = jan,
  journal = {Critical Studies on Terrorism},
  volume = {17},
  number = {1},
  pages = {21--43},
  publisher = {Routledge},
  issn = {1753-9153},
  doi = {10.1080/17539153.2024.2305742},
  urldate = {2026-04-14}
}

@article{antoniakEvaluatingStabilityEmbeddingbased2018,
  title = {Evaluating the {{Stability}} of {{Embedding-based Word Similarities}}},
  author = {Antoniak, Maria and Mimno, David},
  year = 2018,
  month = feb,
  journal = {Transactions of the Association for Computational Linguistics},
  volume = {6},
  pages = {107--119},
  issn = {2307-387X},
  doi = {10.1162/tacl_a_00008},
  urldate = {2025-09-28}
}

@article{broidoScalefreeNetworksAre2019,
  title = {Scale-Free Networks Are Rare},
  author = {Broido, Anna D. and Clauset, Aaron},
  year = 2019,
  month = mar,
  journal = {Nature Communications},
  volume = {10},
  number = {1},
  pages = {1017},
  publisher = {Nature Publishing Group},
  issn = {2041-1723},
  doi = {10.1038/s41467-019-08746-5},
  urldate = {2025-04-02},
  copyright = {2019 The Author(s)},
  langid = {english}
}

@article{cinelliDynamicsOnlineHate2021,
  title = {Dynamics of Online Hate and Misinformation},
  author = {Cinelli, Matteo and Pelicon, Andra{\v z} and Mozeti{\v c}, Igor and Quattrociocchi, Walter and Novak, Petra Kralj and Zollo, Fabiana},
  year = 2021,
  month = nov,
  journal = {Scientific Reports},
  volume = {11},
  number = {1},
  pages = {22083},
  publisher = {Nature Publishing Group},
  issn = {2045-2322},
  doi = {10.1038/s41598-021-01487-w},
  urldate = {2025-02-13},
  copyright = {2021 The Author(s)},
  langid = {english}
}

@article{clausetPowerLawDistributionsEmpirical2009,
  title = {Power-{{Law Distributions}} in {{Empirical Data}}},
  author = {Clauset, Aaron and Shalizi, Cosma Rohilla and Newman, M. E. J.},
  year = 2009,
  month = nov,
  journal = {SIAM Review},
  volume = {51},
  number = {4},
  pages = {661--703},
  publisher = {{Society for Industrial and Applied Mathematics}},
  issn = {0036-1445},
  doi = {10.1137/070710111},
  urldate = {2025-08-26}
}

@article{doddsOusiometricsEssenceMeaning2026,
  title = {Ousiometrics: {{The}} Essence of Meaning Aligns with a Power-Danger-Structure Framework Instead of Valence-Arousal-Dominance},
  shorttitle = {Ousiometrics},
  author = {Dodds, Peter Sheridan and Alshaabi, Thayer and Fudolig, Mikaela Irene and Zimmerman, Julia Witte and Lovato, Juniper and Beaulieu, Shawn and Minot, Joshua R. and Arnold, Michael V. and Reagan, Andrew J. and Danforth, Christopher M.},
  year = 2026,
  month = may,
  journal = {Science Advances},
  volume = {12},
  number = {19},
  pages = {eadr4039},
  publisher = {American Association for the Advancement of Science},
  doi = {10.1126/sciadv.adr4039},
  urldate = {2026-06-05}
}

@article{gallagherGeneralizedWordShift2021,
  title = {Generalized Word Shift Graphs: A Method for Visualizing and Explaining Pairwise Comparisons between Texts},
  shorttitle = {Generalized Word Shift Graphs},
  author = {Gallagher, Ryan J. and Frank, Morgan R. and Mitchell, Lewis and Schwartz, Aaron J. and Reagan, Andrew J. and Danforth, Christopher M. and Dodds, Peter Sheridan},
  year = 2021,
  month = dec,
  journal = {EPJ Data Science},
  volume = {10},
  number = {1},
  pages = {1--29},
  publisher = {SpringerOpen},
  issn = {2193-1127},
  doi = {10.1140/epjds/s13688-021-00260-3},
  urldate = {2025-04-08},
  copyright = {2021 The Author(s)},
  langid = {english}
}

@phdthesis{gothardIncelLexiconDeciphering2021,
  type = {M.{{S}}.},
  title = {The {{Incel Lexicon}}: {{Deciphering}} the {{Emergent Cryptolect}} of a {{Global Misogynistic Community}}},
  shorttitle = {The {{Incel Lexicon}}},
  author = {Gothard, Kelly Caroline},
  year = 2021,
  address = {United States -- Vermont},
  urldate = {2025-12-06},
  copyright = {Database copyright ProQuest LLC; ProQuest does not claim copyright in the individual underlying works.},
  isbn = {979-8-5355-7024-2},
  langid = {english},
  school = {The University of Vermont and State Agricultural College}
}

@inproceedings{hamiltonDiachronicWordEmbeddings2016,
  title = {Diachronic {{Word Embeddings Reveal Statistical Laws}} of {{Semantic Change}}},
  booktitle = {Proceedings of the 54th {{Annual Meeting}} of the {{Association}} for {{Computational Linguistics}} ({{Volume}} 1: {{Long Papers}})},
  author = {Hamilton, William L. and Leskovec, Jure and Jurafsky, Dan},
  editor = {Erk, Katrin and Smith, Noah A.},
  year = 2016,
  month = aug,
  pages = {1489--1501},
  publisher = {Association for Computational Linguistics},
  address = {Berlin, Germany},
  doi = {10.18653/v1/P16-1141},
  urldate = {2025-09-28}
}

@article{hineKekCucksGod2017,
  title = {Kek, {{Cucks}}, and {{God Emperor Trump}}: {{A Measurement Study}} of 4chan's {{Politically Incorrect Forum}} and {{Its Effects}} on the {{Web}}},
  shorttitle = {Kek, {{Cucks}}, and {{God Emperor Trump}}},
  author = {Hine, Gabriel and Onaolapo, Jeremiah and Cristofaro, Emiliano De and Kourtellis, Nicolas and Leontiadis, Ilias and Samaras, Riginos and Stringhini, Gianluca and Blackburn, Jeremy},
  year = 2017,
  month = may,
  journal = {Proceedings of the International AAAI Conference on Web and Social Media},
  volume = {11},
  number = {1},
  pages = {92--101},
  issn = {2334-0770},
  doi = {10.1609/icwsm.v11i1.14893},
  urldate = {2025-10-28},
  copyright = {Copyright (c) 2021 Proceedings of the International AAAI Conference on Web and Social Media},
  langid = {english}
}

@inproceedings{kutuzovDiachronicWordEmbeddings2018,
  title = {Diachronic Word Embeddings and Semantic Shifts: A Survey},
  shorttitle = {Diachronic Word Embeddings and Semantic Shifts},
  booktitle = {Proceedings of the 27th {{International Conference}} on {{Computational Linguistics}}},
  author = {Kutuzov, Andrey and {\O}vrelid, Lilja and Szymanski, Terrence and Velldal, Erik},
  editor = {Bender, Emily M. and Derczynski, Leon and Isabelle, Pierre},
  year = 2018,
  month = aug,
  pages = {1384--1397},
  publisher = {Association for Computational Linguistics},
  address = {Santa Fe, New Mexico, USA},
  urldate = {2026-06-30}
}

@article{mccabePostJanuary6thDeplatforming2024,
  title = {Post-{{January}} 6th Deplatforming Reduced the Reach of Misinformation on {{Twitter}}},
  author = {McCabe, Stefan D. and Ferrari, Diogo and Green, Jon and Lazer, David M. J. and Esterling, Kevin M.},
  year = 2024,
  month = jun,
  journal = {Nature},
  volume = {630},
  number = {8015},
  pages = {132--140},
  publisher = {Nature Publishing Group},
  issn = {1476-4687},
  doi = {10.1038/s41586-024-07524-8},
  urldate = {2025-10-17},
  copyright = {2024 The Author(s), under exclusive licence to Springer Nature Limited},
  langid = {english}
}

@misc{mikolovEfficientEstimationWord2013,
  title = {Efficient {{Estimation}} of {{Word Representations}} in {{Vector Space}}},
  author = {Mikolov, Tomas and Chen, Kai and Corrado, Greg and Dean, Jeffrey},
  year = 2013,
  month = sep,
  number = {arXiv:1301.3781},
  eprint = {1301.3781},
  primaryclass = {cs},
  publisher = {arXiv},
  doi = {10.48550/arXiv.1301.3781},
  urldate = {2025-04-08},
  archiveprefix = {arXiv}
}

@article{newmanMixingPatternsNetworks2003,
  title = {Mixing Patterns in Networks},
  author = {Newman, M. E. J.},
  year = 2003,
  journal = {Physical Review E},
  volume = {67},
  number = {2},
  doi = {10.1103/PhysRevE.67.026126}
}

@article{ngCrossplatformInformationSpread2022,
  title = {Cross-Platform Information Spread during the {{January}} 6th Capitol Riots},
  author = {Ng, Lynnette Hui Xian and Cruickshank, Iain J. and Carley, Kathleen M.},
  year = 2022,
  month = sep,
  journal = {Social Network Analysis and Mining},
  volume = {12},
  number = {1},
  pages = {133},
  issn = {1869-5469},
  doi = {10.1007/s13278-022-00937-1},
  urldate = {2025-10-17},
  langid = {english}
}

@incollection{pancieraWikipediansAreBorn2009,
  title = {Wikipedians Are Born, Not Made},
  booktitle = {Proceedings of the 2009 {{ACM International Conference}} on {{Supporting Group Work}}},
  author = {Panciera, Katherine and Halfaker, Aaron and Terveen, Loren},
  year = 2009,
  month = may,
  series = {{{ACM Conferences}}},
  pages = {51--60},
  doi = {10.1145/1531674.1531682},
  urldate = {2026-06-02},
  isbn = {978-1-60558-500-0}
}

@article{pfefferHalfLifeTweet2023,
  title = {The {{Half-Life}} of a {{Tweet}}},
  author = {Pfeffer, J{\"u}rgen and Matter, Daniel and Sargsyan, Anahit},
  year = 2023,
  month = jun,
  journal = {Proceedings of the International AAAI Conference on Web and Social Media},
  volume = {17},
  pages = {1163--1167},
  issn = {2334-0770},
  doi = {10.1609/icwsm.v17i1.22228},
  urldate = {2026-03-31},
  copyright = {Copyright (c) 2023 Association for the Advancement of Artificial Intelligence},
  langid = {english}
}

@misc{plessKiwiFarmsWebs2016,
  title = {Kiwi {{Farms}}, the {{Web}}'s {{Biggest Community}} of {{Stalkers}}},
  author = {Pless, Margaret},
  year = 2016,
  month = jul,
  journal = {Intelligencer},
  urldate = {2026-02-04},
  howpublished = {https://nymag.com/intelligencer/2016/07/kiwi-farms-the-webs-biggest-community-of-stalkers.html},
  langid = {english}
}

@article{scrivensMeasuringEvolutionRadical2020,
  title = {Measuring the {{Evolution}} of {{Radical Right-Wing Posting Behaviors Online}}},
  author = {Scrivens, Ryan and Davies, Garth and Frank, Richard},
  year = 2020,
  month = feb,
  journal = {Deviant Behavior},
  publisher = {Routledge},
  issn = {0163-9625},
  urldate = {2025-10-28},
  copyright = {\copyright{} 2018 Taylor \& Francis Group, LLC},
  langid = {english}
}

@misc{soniAbolitionistNetworksModeling2021,
  title = {Abolitionist {{Networks}}: {{Modeling Language Change}} in {{Nineteenth-Century Activist Newspapers}}},
  shorttitle = {Abolitionist {{Networks}}},
  author = {Soni, Sandeep and Klein, Lauren and Eisenstein, Jacob},
  year = 2021,
  month = mar,
  journal = {arXiv.org},
  urldate = {2026-07-01},
  howpublished = {https://arxiv.org/abs/2103.07538v1},
  langid = {english}
}

@article{tornbergWhitePowerEcho2024,
  title = {Inside a {{White Power}} Echo Chamber: {{Why}} Fringe Digital Spaces Are Polarizing Politics},
  shorttitle = {Inside a {{White Power}} Echo Chamber},
  author = {T{\"o}rnberg, Petter and T{\"o}rnberg, Anton},
  year = 2024,
  month = aug,
  journal = {New Media \& Society},
  volume = {26},
  number = {8},
  pages = {4511--4533},
  publisher = {SAGE Publications},
  issn = {1461-4448},
  doi = {10.1177/14614448221122915},
  urldate = {2026-04-22},
  langid = {english}
}

@article{tornbergWhiteSupremacistsAnonymous2025,
  title = {White Supremacists Anonymous: How Digital Media Emotionally Energize Far-Right Movements},
  shorttitle = {White Supremacists Anonymous},
  author = {T{\"o}rnberg, Anton and T{\"o}rnberg, Petter},
  year = 2025,
  month = jan,
  journal = {Journal of Information Technology \& Politics},
  volume = {22},
  number = {1},
  pages = {131--148},
  publisher = {Routledge},
  issn = {1933-1681},
  doi = {10.1080/19331681.2023.2262459},
  urldate = {2026-04-22}
}

@article{vanschenckReplatformizationExpansionAlttech2026,
  title = {Replatformization and the Expansion of the Alt-Tech Ecosystem on Kiwi Farms},
  author = {Van Schenck, Reed},
  year = 2026,
  month = aug,
  journal = {Information, Communication \& Society},
  volume = {0},
  number = {0},
  pages = {1--18},
  publisher = {Routledge},
  issn = {1369-118X},
  doi = {10.1080/1369118X.2026.2713669},
  urldate = {2026-08-19}
}

@article{vermaHowUSPresidential2024,
  title = {How {{U}}.{{S}}. {{Presidential}} Elections Strengthen Global Hate Networks},
  author = {Verma, Akshay and Sear, Richard and Johnson, Neil},
  year = 2024,
  month = oct,
  journal = {npj Complexity},
  volume = {1},
  number = {1},
  pages = {1--6},
  publisher = {Nature Publishing Group},
  issn = {2731-8753},
  doi = {10.1038/s44260-024-00018-8},
  urldate = {2025-02-13},
  copyright = {2024 The Author(s)},
  langid = {english}
}

@article{vuExtremeBBDatabaseLargeScale2023,
  title = {{{ExtremeBB}}: {{A Database}} for {{Large-Scale Research}} into {{Online Hate}}, {{Harassment}}, the {{Manosphere}} and {{Extremism}}},
  shorttitle = {{{ExtremeBB}}},
  author = {Vu, Anh and Wilson, Lydia and Chua, Yi Ting and Shumailov, Ilia and Anderson, Ross},
  year = 2023,
  month = jul,
  doi = {10.17863/CAM.97191},
  urldate = {2026-02-24},
  langid = {english}
}

@inproceedings{vuNoEasyWay2024,
  title = {No {{Easy Way Out}}: The {{Effectiveness}} of {{Deplatforming}} an {{Extremist Forum}} to {{Suppress Hate}} and {{Harassment}}},
  shorttitle = {No {{Easy Way Out}}},
  booktitle = {2024 {{IEEE Symposium}} on {{Security}} and {{Privacy}} ({{SP}})},
  author = {Vu, Anh V. and Hutchings, Alice and Anderson, Ross},
  year = 2024,
  month = may,
  pages = {717--734},
  issn = {2375-1207},
  doi = {10.1109/SP54263.2024.00007},
  urldate = {2026-03-04}
}

@article{zhengAdaptiveLinkDynamics2024,
  title = {Adaptive Link Dynamics Drive Online Hate Networks and Their Mainstream Influence},
  author = {Zheng, Minzhang and Sear, Richard F. and Illari, Lucia and Restrepo, Nicholas J. and Johnson, Neil F.},
  year = 2024,
  month = apr,
  journal = {npj Complexity},
  volume = {1},
  number = {1},
  pages = {1--8},
  publisher = {Nature Publishing Group},
  issn = {2731-8753},
  doi = {10.1038/s44260-024-00002-2},
  urldate = {2025-02-13},
  copyright = {2024 The Author(s)},
  langid = {english}
}

\section{Supplementary tables and figures}\label{secA1}


\begin{table}[htbp]
\begin{tabular}{lrr}
\toprule
power law vs. & loglikelihood ratio & significance \\
\midrule
lognormal & 0.21 & 0.30 \\
exponential & 31649.5 & $4.52\times 10^{-126}$ \\
stretched exponential & 221.26 & $2.94\times 10^{-30}$ \\
\bottomrule
\end{tabular}
\begin{tabular}{lrr}
\toprule
power law vs. & loglikelihood ratio & significance \\
\midrule
lognormal & 0.0046 & 0.87 \\
exponential & 33.37 & $1.34\times 10^{-5}$ \\
stretched exponential & 1.19 & 0.13 \\
\bottomrule
\end{tabular}
\caption{Fit comparisons between power law and three component distributions for the thread length (top) and degree in the thread hyperlink network (bottom). We note that while a power law is the best fit of these distributions in both cases, its improvement over lognormal is not statistically significant for thread-length distribution, nor is the improvement over lognormal or exponential for the degree distribution.}
\label{tab:dist-comparisons}
\end{table}

\begin{figure}[h]
    \centering
    \includegraphics[height=.3\textheight]{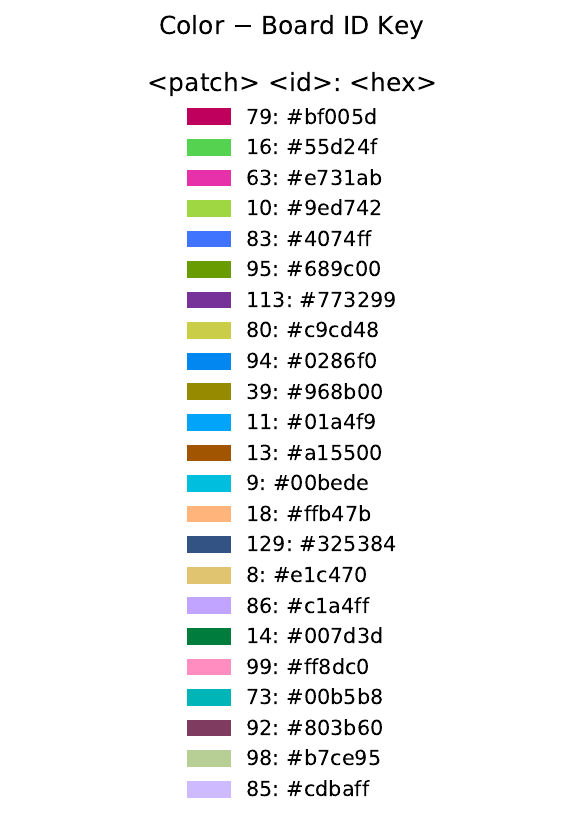}
    \caption{The legend indicating the board ID number for each color represented in Figure \ref{fig:internal-network}. Each number corresponds to a board where threads with the same overarching topic are situated. Some example board descriptions are given in Section \ref{sec:temp-link-network}.}
    \label{fig:network-legend}
\end{figure}

\begin{figure}[h]
    \centering
    \includegraphics[width=.9\textwidth]{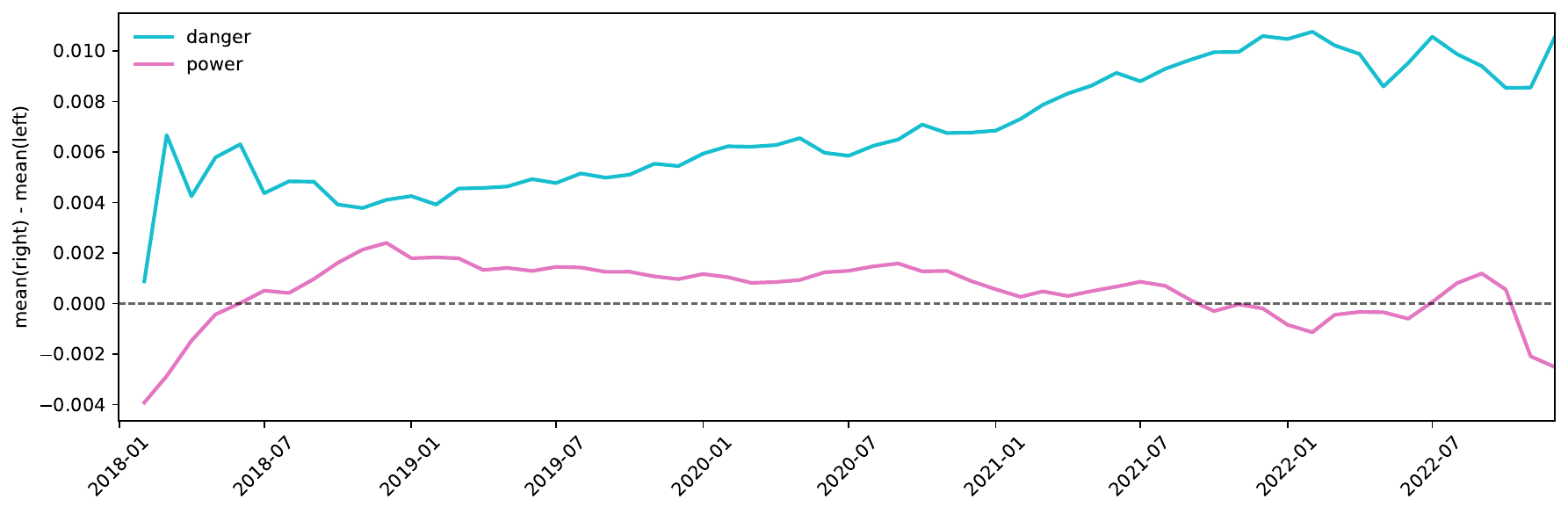}
    \caption{The difference in danger-similarity and power-similarity between the left and right regions as a function of the date chosen for the partition between the regions.}
    \label{fig:sliding-breakpoint}
\end{figure}

\begin{figure}[h]
    \centering 
    \includegraphics[width=.9\textwidth]{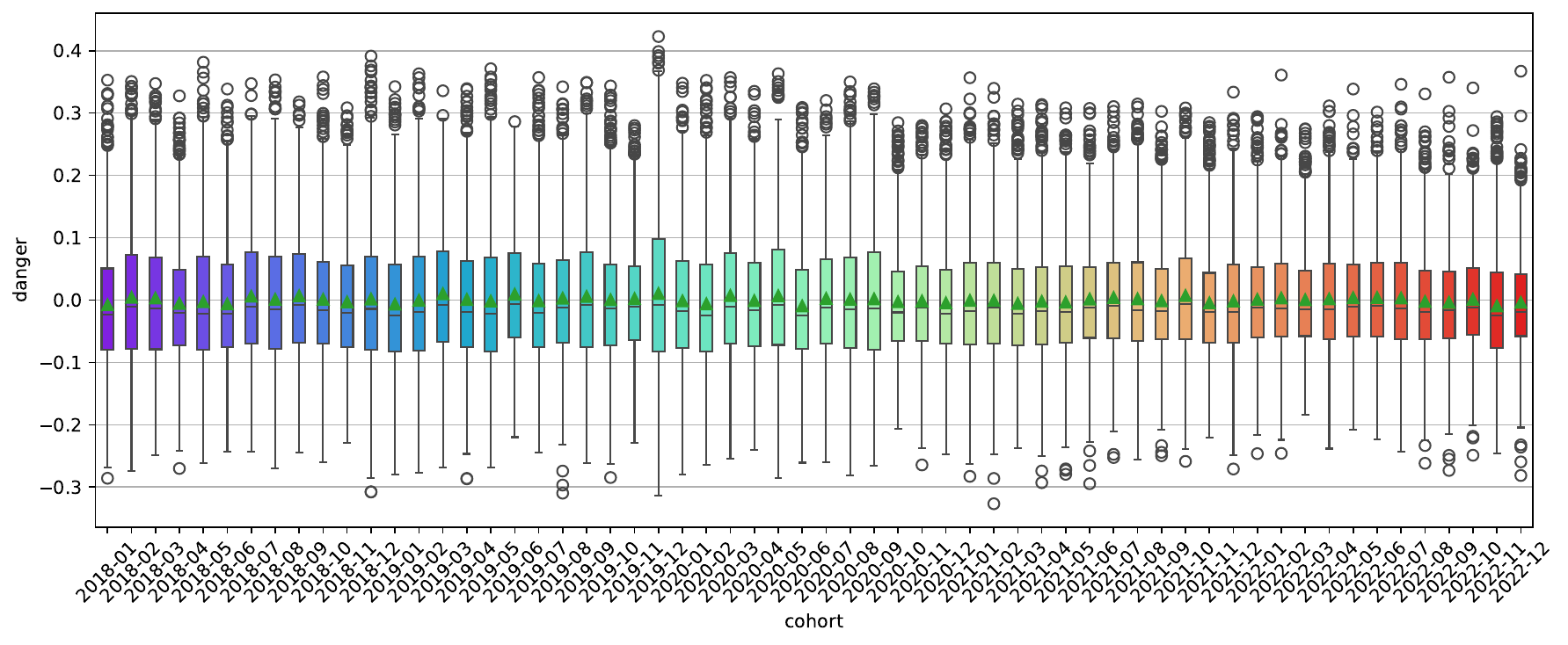}    
    \caption{Danger-similarity distribution boxplots of most similar words to ``lolcow'' for each monthly cohort, over their tenures through 2022. Box edges and horizontal lines indicate quartiles, and triangles indicate means.}
    \label{fig:embed-monthly-cohorts}
\end{figure}

\end{appendices}

\end{document}